\documentclass[11pt,floatfix,amssymb,pra,aps]{revtex4-1}

\newcommand{\xzpf}{x_{\mathrm{zpf}}}
\newcommand{\pzpf}{p_{\mathrm{zpf}}}
\newcommand{\XC}{X_{\scriptscriptstyle{C}}}
\newcommand{\YC}{Y_{\scriptscriptstyle{C}}}

\newcommand{\nm}{\bar{n}}
\newcommand{\neff}{{n}_{\mathrm{eff}}}
\newcommand{\nmin}{{n}_{\mathrm{min}}}
\newcommand{\nmax}{{n}_{\mathrm{max}}}

\newcommand{\Gm}{\Gamma_{\mathrm{m}}}

\newcommand{\GY}{\Gamma_{\scriptscriptstyle{Y}}}
\newcommand{\GX}{\Gamma_{\scriptscriptstyle{X}}}

\newcommand{\caXY}{a_{\scriptscriptstyle{X,Y}}}

\newcommand{\caX}{a_{\scriptscriptstyle{X}}}
\newcommand{\cbX}{b_{\scriptscriptstyle{X}}}
\newcommand{\caY}{a_{\scriptscriptstyle{Y}}}
\newcommand{\cbY}{b_{\scriptscriptstyle{Y}}}

\newcommand{\Gnj}{\Gamma_{\mathrm{n}j}}

\newcommand{\GdsX}{\Gamma^{\mathrm{d}}_{\scriptscriptstyle{X}}}
\newcommand{\GdsY}{\Gamma^{\mathrm{d}}_{\scriptscriptstyle{Y}}}

\newcommand{\Omegamj}{\Omega_{j}}
\newcommand{\OmegaX}{\Omega_{\scriptscriptstyle{X}}}
\newcommand{\OmegaY}{\Omega_{\scriptscriptstyle{Y}}}
\newcommand{\omegaL}{\omega_{\scriptscriptstyle{\mathrm{L}}}}
\newcommand{\gX}{g_{\scriptscriptstyle{X}}}
\newcommand{\gY}{g_{\scriptscriptstyle{Y}}}
\newcommand{\bX}{\hat{b}_{\scriptscriptstyle{X}}}

\newcommand{\bdX}{\hat{b}^{\dagger}_{\scriptscriptstyle{X}}}

\newcommand{\ac}{\hat{a}_{\mathrm{c}}}

\newcommand{\kB}{k_{\mathrm{B}}}

\newcommand{\ud}{\mathrm{d}}

\usepackage{amsmath}
\usepackage[dvips]{graphicx}
\usepackage{epsfig}
\usepackage{tabularx}
\usepackage{color}
\begin{document}

\title{Two-dimensional quantum motion of a levitated nanosphere}

\author{A. Ranfagni}
\affiliation{European Laboratory for Non-Linear Spectroscopy (LENS), via Carrara 1, I-50019 Sesto Fiorentino (FI), Italy}
\affiliation{INFN, Sezione di Firenze, via Sansone 1, I-50019 Sesto Fiorentino (FI), Italy}

\author{K. B\o rkje}
\affiliation{Department of Science and Industry Systems, University of South-Eastern Norway, PO Box 235, NO-3603 Kongsberg, Norway}

\author{F. Marino}
\affiliation{CNR-INO, largo Enrico Fermi 6, I-50125 Firenze, Italy}
\affiliation{INFN, Sezione di Firenze, via Sansone 1, I-50019 Sesto Fiorentino (FI), Italy}

\author{F. Marin}
\email[Electronic mail: ]{marin@fi.infn.it}
\affiliation{Dipartimento di Fisica e Astronomia, Universit\`a di Firenze, via Sansone 1, I-50019 Sesto Fiorentino (FI), Italy}
\affiliation{European Laboratory for Non-Linear Spectroscopy (LENS), via Carrara 1, I-50019 Sesto Fiorentino (FI), Italy}
\affiliation{INFN, Sezione di Firenze, via Sansone 1, I-50019 Sesto Fiorentino (FI), Italy}
\affiliation{CNR-INO, largo Enrico Fermi 6, I-50125 Firenze, Italy}

\date{\today}

\begin{abstract}
We report on the two-dimensional (2D) dynamics of a levitated nanoparticle in an optical cavity. The motion of the nanosphere is strongly coupled to the cavity field by coherent scattering and heavily cooled in the plane orthogonal to the tweezer axis.  Due to the characteristics of the 2D motion and the
strong optomechanical coupling, the motional sideband asymmetry that reveals the quantum nature of the dynamics is not limited to mere scale factors between Stokes and anti-Stokes peaks, as customary in quantum optomechanics, but assumes a peculiar spectral dependence. We introduce and discuss an effective thermal occupancy that quantifies how close the system is to a minimum uncertainty state and allows us to consistently characterize the particle motion. By rotating the polarization angle of the tweezer beam we tune the system from a one-dimensional (1D) cooling regime, where we achieve a best thermal occupancy of $0.51 \pm 0.05$, to a regime in which the fully 2D dynamics of the particle exhibits strong non-classical properties. We achieve a strong 2D confinement with thermal occupancy of $3.4 \pm 0.4$ along the warmest direction and around unity in the orthogonal one.
These results represents a major improvement with respect to previous experiments both considering the 1D and 2D motion, and pave the way towards the preparation of tripartite optomechanical entangled states and novel applications to directional force and displacement quantum sensing.
\end{abstract}

\maketitle

Trapping and levitation of micrometric objects in strongly focused light beams (optical tweezers) was introduced in the pioneering works by Arthur Ashkin in the early 1970s \cite{Ashkin:1970}, and the major impact of such techniques in a multidisciplinary environment earned him the Nobel Prize in Physics 2018. In most applications, optical trapping occurs in a strongly damping background.  The idea of operating in a high vacuum, thus reducing the interaction with the environment, and bringing levitating and oscillating nano-particles into the quantum regime was boosted about 10 years ago \cite{Chang:2010,Barker:2010,Romero:2010,Li:2011} and has since developed into a fruitful research topic \cite{Millen2020}. The goal of cooling the oscillations, at least along one direction, down to a phononic occupation number below unity, has been recently achieved. By means of electrostatic feedback cooling on charged particles the motion was frozen along the tweezer propagation axis \cite{Magrini2021,Tebbenjohanns2021}, while cavity cooling produced similar results for the oscillation along an axis in the transverse plane \cite{delic2020}. Cavity cooling, initially implemented with standard optomechanical methods \cite{Kiesel:2013}, has become much more efficient with the introduction of the coherent scattering technique \cite{delic2019,windey2019,gonzalez-ballestrero2019} imported from atomic physics experiments \cite{Vuletic2000}. 
In this configuration, the cavity is driven by tweezer light scattered by the trapped particle. As in standard sideband cooling experiments, an optical field red-detuned with respect to the cavity resonance produces cold damping of the particle motion.
However, since the cavity mode is only populated by photons coherently scattered by the particle, such a scheme allows for a higher optomechanical coupling between optical and mechanical modes, by circumventing the power limitations associated with the usual dispersive coupling. 

Optically levitating nanoparticles in high vacuum offer a quite natural platform for the study of quantum mechanical features in all three spatial dimensions and the achievement of quantum coherent control of their motion, with applications ranging from quantum foundations and information processing to directional quantum sensing. 

The optical potential experienced by the nano-particle is proportional to the light intensity. As a consequence, its oscillatory motion is characterized by a lower frequency along the tweezer propagation axis, where the characteristic length is the Rayleigh range, with respect to the transverse plane, where the characteristic length is the beam waist. For linearly polarized light, the strongly focused beam is slightly elliptical, with the largest diameter along the electric field \cite{Novotny:2012}, producing different transverse oscillation frequencies. Consequently, while the motion along the tweezer propagation axis is strongly decoupled from the others and can be treated as a single harmonic oscillator, on the transverse plane the system appears as truly two-dimensional (2D), in particular when the motion of the nanosphere is coupled with the modes of an optical cavity. In this case, indeed, it cannot be simply decomposed into two independent oscillator modes, due to their mutual interaction mediated by the optical field. 

An important step forward on the route towards the ambitious goal of realizing three-dimensional ground state cooling is the observation and characterization of the 2D quantum motion on the tweezer transverse plane. 
An optomechanical system with two nearly-degenerate mechanical modes was already considered in the literature \cite{genes2008,massel2012,shkarin}. However, here we deal with 2D motion, where its projections along all the directions (i.e., every linear combination of the two modes that are arbitrarily chosen as reference frame) have a clear physical meaning. 
This requires addressing the problem in a radically different way from what is customary in optomechanics, both in the description of the system and in the analysis of the experimental signals and of the information that can be extracted.
The problem is theoretically analyzed in \cite{toros2021} where it is shown that for suitable system parameters the particle full planar motion can be strongly coupled to the light and efficiently cooled through coherent scattering. Recently, the quantum-coherent strong coupling between the 2D motion of a levitating nanoparticle and the radiation field has been demonstrated, and led to the observation of vectorial optomechanical polaritons \cite{Ranfagni2021}. Nevertheless, the ground state of the 2D motion has never been approached so far.

In this work we exploit coherent scattering to cool the motion of a nanosphere in the plane orthogonal to the propagation axis of the tweezer, in the regime of strong optomechanical coupling with the cavity field. We go deeper into the quantum regime than in previous works, both by considering the motion in the coldest direction, and by considering two-dimensional motion. We show how the spectrum of the electromagnetic field exiting the optical cavity displays a very strong non-classical signature, but it is not correct to directly deduce a phononic occupation number from the measured peak asymmetry as in standard sideband thermometry \cite{Safavi2012,Khalili2012,Borkje:2016}. The characteristics of the motion are instead obtained from a more complete analysis of the spectra. The comparison of the experimental data with the theoretical modelling allows to derive the essential system parameters and then reconstruct the spectrum of the motion. In order to quantify the cooling along any direction, we measure how far we are from a minimum uncertainty state. The introduced indicator can be traced back to the usual phononic occupation number in the case of a single oscillator, but it additionally allows to quantify the motion in each direction without inappropriately associating it to a single harmonic oscillator.

\section{Physical background and model}

\begin{figure}
    \centering
    \includegraphics[width=0.9\columnwidth]{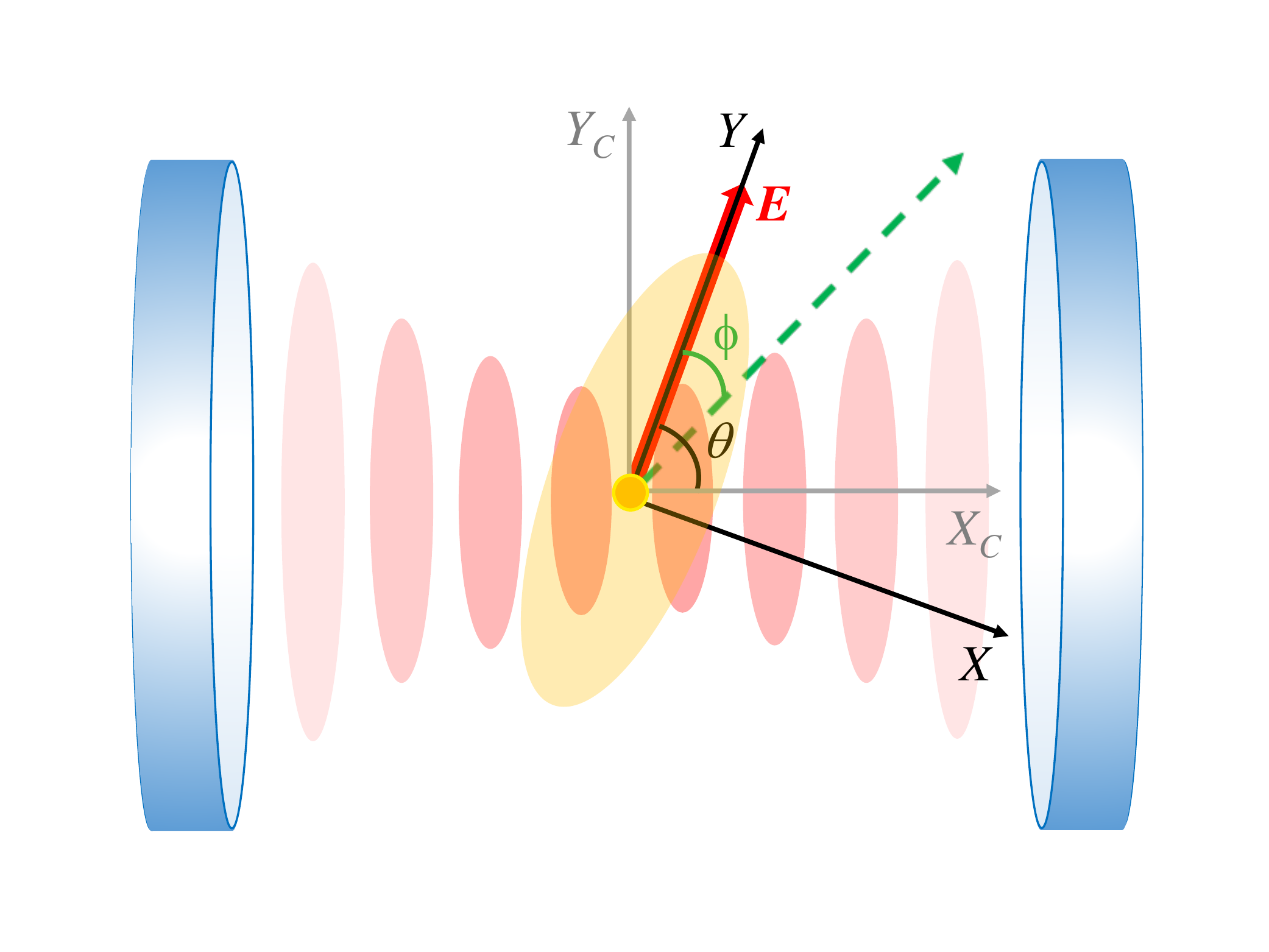}
    \caption{Scheme of the experimental system. A silica nanosphere levitates on the optical potential created by the light of an optical tweezer, propagating along the $Z$ axis, with the electric field along the $Y$ axis. The nanosphere is positioned on the axis of an optical cavity (which defines the direction $\XC$), in correspondence of a node of the field of a cavity mode. The angle $\phi$ defines a generic direction in the transverse plane $X - Y$.   
     }
    \label{fig_setup}
\end{figure}

In the experimental configuration that exploits coherent scattering, the particle is positioned on the optical axis of the cavity, typically with the propagation axis of the tweezer perpendicular to the cavity axis (Fig. \ref{fig_setup}). The optomechanical coupling occurs with the motion along the cavity axis ($\XC$ in Fig. \ref{fig_setup}), and it is optimized by positioning the particle in a node of the standing wave associated with the cavity mode. On the other hand, the coupling is null for the motion orthogonal to the plane formed by the tweezer and the cavity axis ($\YC$ in Fig. \ref{fig_setup}). 

In the case of very weak optomechanical coupling or when the tweezer radiation is far from resonance, the motion in the transverse plane is better described in the framework defined by the polarization axis (we define as $X$ and $Y$ the directions respectively orthogonal and parallel to the electric field). However, if the $X$ axis is not perfectly coincident with the cavity axis (and consequently the motion along $Y$ is not perfectly decoupled), a correct calculation of the optomechanical effects on the $X$ mode, and in particular of its effective temperature, requires to consider its indirect coupling (mediated by the electromagnetic field) with the $Y$ mode \cite{genes2008,toros2020}. The latter, being weakly coupled with the field, remains relatively hot. 

The picture is different in case of optimal detuning ($\Delta \simeq -0.5(\OmegaX+\OmegaY)$; here $\Delta=\omegaL-\omega_{\mathrm{c}}$ is the detuning of the tweezer radiation at frequency $\omegaL/2\pi$ with respect to the cavity resonance frequency $\omega_{\mathrm{c}}/2\pi$, and $\Omega_{\scriptscriptstyle{X,Y}}/2\pi$ are the mechanical resonance frequencies) and sufficiently strong coupling $g$ ($4g^2/\kappa \gtrsim | \OmegaX-\OmegaY |$, where $\kappa$ is the optical decay rate). Now, the optomechanical coupling is more important than the frequency difference between the modes in determining the system's dynamical behavior, and it is more appropriate to use the cavity axis $\XC$ and $\YC$ as framework for describing the 2D motion. The two directions define respectively the so-called geometrical bright and the dark modes \cite{toros2021}. Again, however, unless the mechanical frequencies $\Omega_{\scriptscriptstyle{X,Y}}$ are perfectly degenerate, bright and dark modes are coupled together. The dark mode is sympathetically cooled by the bright mode, which is in turn warmed by the dark mode. 

In any case, as long as $g < \kappa/4$ (weak optomechanical coupling regime), the motions along any direction, including all the four axes ($X$, $Y$, $\XC$, $\YC$), have spectra characterized by two peaks, centred on the eigenfrequencies of the optomechanical system. It is therefore not straightforward to associate a single harmonic oscillator to any of these directions, as it could be done for the directions corresponding to the eigenstates of the drift matrix. In the case of strong optomechanical coupling ($g > \kappa/4$), the mechanical modes hybridize with the optical one, and it is even less obvious how to associate a harmonic oscillator with the motion in any direction and assign phononic occupation numbers as usual in optomechanics. In Section \ref{one-dimensional} we will introduce a rigorous procedure to define an effective thermal occupancy for the motion in each direction.

For the purpose of reproducing the experimental results, the oscillation of the nanosphere in the optical potential created by the tweezer, and its interaction with the mode of the cavity field that is close to resonance with the tweezer light, can be described by a standard quantum Langevin model. The linearized evolution equations for the motion in the plane orthogonal to the tweezer axis, expressed in the frame rotating at the laser frequency $\omegaL$, can be written as 
\begin{equation}
\dot{\hat{a}}_{\mathrm{c}}=\bigg(  i\Delta-\frac{\kappa}{2}\bigg)\ac+i\gX(\hat{b}_X+\hat{b}_X^\dag)+i\gY(\hat{b}_Y+\hat{b}_Y^\dag)+\sqrt{\kappa}\,\hat{a}_{\mathrm{in}}
\label{dta}
\end{equation}
\begin{equation}
\dot{\hat{b}}_j=\left(-i\Omegamj-\frac{\Gm}{2}\right)\hat{b}_j+ig_j(\ac+\ac^{\dag})+\sqrt{\Gamma_j}\,\hat{b}_{\mathrm{n},j} 
\label{dtb}
\end{equation}
where the operators $\ac$ and $\hat{b}_j$ describe respectively the intracavity field and the two mechanical modes ($j=X, Y$), $\Gm$ is the gas damping rate, and $g_j$ are the optomechanical coupling rates. The $\hat{b}_j$ are linked to the operators describing the displacements ($x$, $y$) and the momenta ($p_x$, $p_y$) along the $X$ and $Y$ directions by the relations
\begin{eqnarray}
\label{eq_xpx}
x &=& \xzpf (\bX+\bdX) \\
p_x &=& \pzpf\, i (\bdX-\bX)
\end{eqnarray}
where $\,\xzpf = \sqrt{\frac{\hbar}{2 m \OmegaX}}$ and $\,\pzpf = \sqrt{\frac{\hbar m \OmegaX}{2 }}$ are the zero-point position and momentum fluctuations of the free oscillators, and by equivalent expressions for $y$ and $p_y$. We note that while the operators $\hat{b}_j$ can be thought of as phonon annihilation operators in the limit of weak optomechanical coupling, such an interpretation is not straightforward in the strong coupling regime.

The input noise operators are characterized by the correlation functions
\begin{eqnarray}
\langle\hat{a}_{\mathrm{in}}(t)\hat{a}_{\mathrm{in}}^{\dag}(t')\rangle & = & \delta(t-t')
\label{noise1} \\
\langle\hat{a}_{\mathrm{in}}^{\dag}(t)\hat{a}_{\mathrm{in}}(t')\rangle & = & 0
\label{noise2}  \\
\langle\hat{b}_{\mathrm{n},j}^{\dag}(t)\hat{b}_{\mathrm{n},j}(t')\rangle = \langle\hat{b}_{\mathrm{n},j}(t)\hat{b}_{\mathrm{n},j}^{\dag}(t')\rangle & = & \delta(t-t')
\label{noise5}
\end{eqnarray}
The total decoherence rates can be written as 
\begin{equation}
\label{eq_Gamma}
%\Gamma_j = \bar{n}_{\mathrm{th},j} \Gm + \Gnj
\Gamma_j = \frac{\kB T}{\hbar \Omegamj}\, \Gm + \Gnj
\end{equation}
where the first contribution is due to scattering with the background gas molecules at temperature $T$,  
and $\Gnj$ accounts for the shot noise in the dipole scattering (which adds negligible damping), as well as for additional technical noise sources. We note that in Eq. (\ref{noise5}) we are dealing with noise in a classical way, justified by the fact that $\kB T \gg \hbar \Omegamj$. 

As discussed, the $\XC$-$\YC$ framework can be more appropriate to understand the physics of the system and to describe it by means of approximated analytical expressions. However, we remark that in the $X$-$Y$ framework the input noise sources are uncorrelated and easily evaluated, and it is therefore more suitable for exact numerical calculations. 

The total cavity output field is given by the input-output relation $\hat{a} = \sqrt{\kappa} \hat{a}_{\mathrm{c}}-\hat{a}_{\mathrm{in}}$. We have solved the above equations in the Fourier space. The heterodyne spectrum normalized to shot noise is
\begin{equation}
S_{\mathrm{out}}(\Omega) = \Big(S_{a a}(\Omega-\Omega_{\mathrm{LO}})+S_{a^{\dagger} a^{\dagger}}(\Omega+\Omega_{\mathrm{LO}})\Big) \eta+ (1-\eta)
\label{eq_Sout}
\end{equation}
where $S_{a a} (\Omega) = \frac{1}{2\pi}\int \ud \Omega ' \langle \tilde{a}^{\dagger}(\Omega ') \tilde{a}(\Omega)\rangle$, $\Omega_{\mathrm{LO}}$ is the angular frequency of the local oscillator, and $\eta$ is the detection efficiency. 
As described in the following sections, the spectrum of Eq. (\ref{eq_Sout}) is fitted to the experimental heterodyne spectra to derive the system parameters in different conditions. The model is then used to calculate the effective thermal occupancy, by integrating numerical spectra as detailed in Section \ref{one-dimensional}.  

\section{Experimental setup}

The silica nanospheres used for this work have a mean diameter of $125 \pm 5\,$nm, measured by photon correlation imaging \cite{Duri2009}. The experimental procedure starts by loading a nanosphere in a first vacuum chamber, on a first optical tweezer mounted on the tip of a movable rod \cite{mestres2015,calamai2020}. The trapped particle is then translated to the experimental chamber, where it is transferred to a second optical tweezer, created by the light of a Nd:YAG laser delivered in vacuum by a polarization maintaining optical fiber \cite{calamai2020}. About 300 mW are focused to an elliptical shape of waists $\sim 1.0 \mu$m and $\sim 0.9 \mu$m, where the tighter focus occurs in the direction orthogonal to the axis defined by the linear polarization of the tweezer light \cite{Novotny:2012}. The corresponding typical oscillation frequencies of the nanosphere in the tweezer optical potential are  27 kHz ($Z$ direction, along the tweezer axis), 115 kHz ($Y$ direction, along the polarization axis) and 128 kHz ($X$ direction). After the transfer, the movable rod is retracted, the two chambers are isolated and the experimental chamber is evacuated to high vacuum.

The second tweezer is mounted on a three-axes nanometric motorized linear stage that allows to position the nanosphere inside a nearly-concentric Fabry-Perot cavity (free-spectral-range $FSR = 3.07$ GHz, linewidth $\kappa/2\pi = 57$ kHz), almost orthogonal to the tweezer axis (Fig. \ref{fig_setup}). A second, auxiliary Nd:YAG laser is frequency locked to the optical cavity, and the tweezer laser is phase locked to the auxiliary one with a controllable frequency offset equal to $FSR + \Delta/2\pi$, thus accurately defining the detuning $\Delta$ of the tweezer radiation from the cavity resonance. 

The light scattered by the particle on the almost resonant cavity mode, and transmitted by the output mirror, is superimposed on a local oscillator beam, derived from the main Nd:YAG laser before launching it in the fiber, and frequency shifted by 0.9 MHz. The mixed beams are sent to a balanced detection to implement a heterodyne measurement. The beat note between the local oscillator and the scattered light allows to deduce the location of the nanosphere inside the mode standing wave, and to place it on the optical axes, in correspondence of a node. In this position, the optical coupling between the particle motion and the radiation field is due to the coherent scattering of the tweezer light on the cavity mode and it is just effective for the projection of the motion on the cavity axis \cite{delic2019,windey2019,gonzalez-ballestrero2019,toros2020}. More details on the experimental setup can be found in Ref. \cite{Ranfagni2021} and its Supplemental Information. 

\section{Measurement of the decoherence rates}

The force noise that determines the motion of the nanosphere can be ascribed to two fundamental sources, besides the optomechanical coupling: the collisions with the molecules of background gas, and the recoil due to the shot-noise in the dipole emission of the nanosphere. In the regime of interest (high vacuum, where the mean free path of the gas molecules is much larger than the diameter of the nanosphere), the expected collisional damping rate is proportional to the pressure $ P $ in the experimental chamber, and is given by \cite{Epstein,Beresnev:1990}
\begin{equation}
\Gm\,=\,\frac{8 \sqrt{\pi}}{3} \frac{R^2}{m}\sqrt{\frac{m_{\mathrm{gas}}}{2 k_{\mathrm{B}} T}}\left(2+\frac{\pi}{4}\right)\,P 
\label{Eq_G_vs_p}
\end{equation}
where $R$ is the radius of the nanosphere, $m$ is its mass, and $m_{\mathrm{gas}}$ is the mass of the gas molecules.
At very low pressures, the contribution of dipole scattering becomes important. In case of linear polarization of the optical tweezer radiation, the theoretical expressions for the decoherence rate due to shot noise are \cite{Seberson2020}
\begin{equation}
\GdsY\,=\,\frac{1}{5} \frac{\hbar \omegaL^2}{2 m c^2 \OmegaY} \frac{I_{\mathrm{tw}}\, \sigma}{\hbar \omegaL}
\label{Eq_GdsY}
\end{equation}
along the polarization axis, and $\GdsX\,=\,2 \GdsY \frac{\OmegaY}{\OmegaX}$
along the $X$ direction (perpendicular to both the polarization axis and the tweezer propagation axis). Here, $I_{\mathrm{tw}}$ is the tweezer intensity at the nanosphere position, and $\sigma$ is the scattering cross section. Jain \emph{et al.} \cite{Jain2016} have observed the re-heating of the $Y$ mode, after parametric feedback cooling down to a phononic occupation number of $\sim 60$ in high vacuum. Their measured rate agrees with the theory within $10 - 30\%$. 

\begin{figure}
    \centering
    \includegraphics{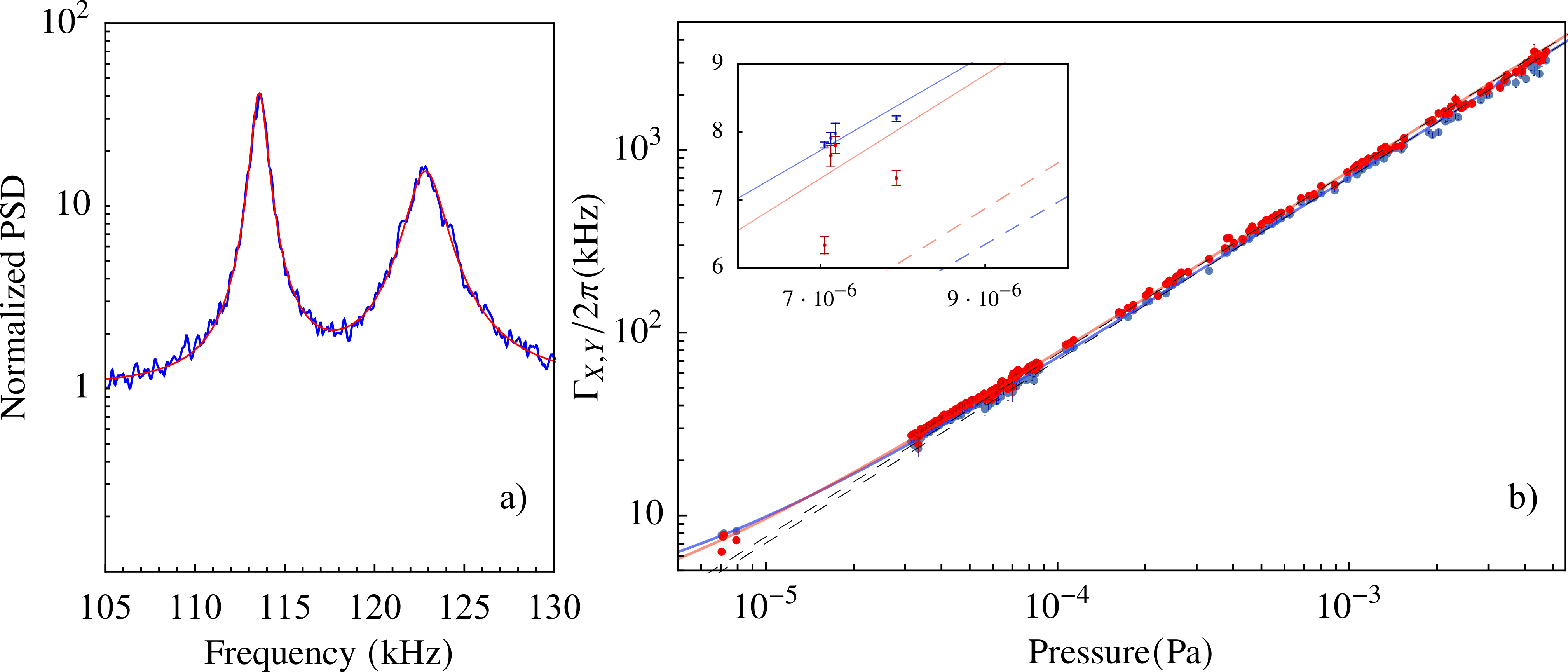}
    \caption{(a) Typical heterodyne spectrum (anti-Stokes sideband, power spectral density PSD normalized to shot noise) showing the resonances of the $X$ and $Y$ modes, acquired at the pressure of $P \simeq 3 \times 10^{-5}\,$Pa for a detuning of $-\Delta / 2 \pi = 220\, \mathrm{kHz} $. The red solid line is the fit with the theoretical model. (b) Total decoherence rates for the $X$ and $Y$ motion (blue and red dots respectively), measured for a tweezer light detuning of $-\Delta / 2 \pi = 220\, \mathrm{kHz} $ as a function of background pressure. The error bars represent the standard error, calculated on several consecutive time intervals. 
    The solid lines show linear fits. The inset is an enlarged view of the region at lowest pressure.  
     }
    \label{fig_Gamma_vs_p}
\end{figure}

We can deduce the total decoherence rates for the $X$ and $Y$ motion from the spectrum of the field transmitted by the cavity. These rates mainly determine the area of the spectral peaks generated by the motion of the nanosphere, while their width is dominated by the optical cooling. We have acquired the time series of the heterodyne detection signal during the evacuation of the experimental chamber, maintaining a relatively large detuning of the tweezer radiation from cavity resonance (namely, $\Delta/2\pi = -220\,$kHz). An example of a derived spectrum (anti-Stokes motional sideband) is shown in Fig. (\ref{fig_Gamma_vs_p}a). The spectra are calculated by Fourier transforming consecutive time intervals, and are fitted to the theoretical model of Eq. (\ref{eq_Sout}). For each spectrum the resonance frequencies  $\OmegaX$ and $\OmegaY$, the optomechanical coupling rates $\gX$ and $\gY$, and the decoherence rates $\GX$ and $\GY$ are derived from the fits, while the detuning $\Delta$, the cavity width $\kappa$ and the detection efficiency $\eta$ are measured independently and kept as fixed parameters in the fitting procedure. 

The decoherence rates vs pressure can be fitted by a straight line, according to  $\Gamma_j /2\pi\,= a_j \,+ b_j\,P$ (Fig. \ref{fig_Gamma_vs_p}b). The parameters derived from this fitting procedure are 
slopes $\cbX\,=\,(7.05 \pm 0.02)\times 10^8\, \mathrm{Hz}/\mathrm{Pa}$ and $\cbY \,=\,(7.64 \pm 0.03)\times 10^8\, \mathrm{Hz}/\mathrm{Pa}$, and constant 
terms $\caX\,=\, 2.79\pm0.06 \, \mathrm{kHz}$ and $\caY\,=\,1.97\pm0.15 \, \mathrm{kHz}$. In these expressions, the quoted errors represent 
the statistical uncertainty of the fits (one standard deviation). However, the main error is due to the evaluation of the detection 
efficiency, measured to be $\eta\,=\,0.295 \pm 0.03$, which produces an additional uncertainty of $10\%$ in the $a_j$ and $b_j$ parameters. 

We have calculated the gas damping rate from the slopes of $\Gamma_j$, according to $\Gm\,=\,P\, b_j\,\frac{\hbar \Omegamj}{\kB T}$ (see Eq. (\ref{eq_Gamma})), obtaining $\Gm/2\pi P\,=\,14.4\,$Hz/Pa for the $X$ mode, and $14.0\,$Hz/Pa for the $Y$ mode, in good relative agreement. With the parameters of our nanosphere in nitrogen atmosphere, the theoretical value given by the expression (\ref{Eq_G_vs_p}) is $\Gm/2\pi P\,=\,9.7\,$Hz/Pa. Even considering the $30\%$ nominal accuracy of the pressure gauge quoted by its manufacturer, the agreement is not very good. This discrepancy can be explained by the heating of the sphere caused by laser absorption, which eventually warms up the background gas, as pointed out in Ref. \cite{Millen2014} and analyzed in Ref. \cite{Hebestreit2018}. 

The decoherence rates due to the photon recoil in the dipole radiation, calculated for our nanosphere using expression (\ref{Eq_GdsY}), are $\GdsX/2\pi\,=\,3.0\,$kHz and $\GdsY/2\pi\,=\,1.7\,$kHz, with a $20\%$ uncertainty derived from the knowledge of the nanosphere mass and of the tweezer intensity. The agreement with the fitted constant terms $\caXY$ is excellent, indicating that no significant extra noise is present, i.e., the $\Gnj$ introduced in Eq. (\ref{eq_Gamma}) coincide with the $\Gamma^{\mathrm{d}}_j$. The calculated fluctuations due to the laser intensity and frequency noise are indeed negligible \cite{Ranfagni2021}, and the measured decoherence rate shows that even parametric heating due to mechanical vibrations, whose relevance was pointed out in other works \cite{windey2019,gonzalez-ballestrero2019}, do not play an evident role in our case. We also remark that we can distinguish the dipole scattering rates in the two modes, whose ratio (weakly dependent on the detection efficiency) we find to be $\caX/\caY\,=\,1.42 \pm 0.14$. To our knowledge, this is the first time that such spatial variation of the shot noise in the photon recoil due to dipole scattering is shown in the motion of a mesoscopic object. The experimental measurement of the photon recoil is the main result described in this Section. The residual discrepancy with the calculated ratio of $1.8$ can be attributed to the imperfectly linear polarization of the light at the output of the optical fiber. 

As a further check of the modeling, we focus on the optomechanical coupling rates. Their theoretical expressions can be written as \cite{delic2019,windey2019} $\gX = g_{\mathrm{max}}\,\sin^2 \theta$ 
and $\gY = g_{\mathrm{max}}\,\sqrt{\frac{\OmegaX}{\OmegaY}}\sin \theta \cos \theta$, 
where $\theta$ is the angle between the cavity axis and the tweezer polarization axis (see Fig. (\ref{fig_setup})), $g_{\mathrm{max}} = \frac{\alpha \epsilon_c \epsilon_{\mathrm{tw}} \omega_c}{2 \hbar c}\sqrt{\frac{\hbar}{2 m \OmegaX}}$ 
($\alpha$ is the nanosphere polarizability), 
$\epsilon_c = \sqrt{\frac{\hbar \omega_c}{2 \epsilon_0 V_c}}$ 
($\epsilon_0$ is the vacuum permittivity, and $V_c$ is the cavity volume), 
and $\epsilon_{\mathrm{tw}} = \sqrt{\frac{2 I_{\mathrm{tw}}}{\epsilon_0 c}}$. 
From the measured optomechanical gains of the two modes, we derive a maximum gain $g_{\mathrm{max}}/2 \pi\,=\,25.7 \pm 1.7\,$kHz, and an angle $\theta$ that varies between $78^\circ$ and $69^\circ$ during the chamber evacuation. The theoretical $g_{\mathrm{max}}/2 \pi$ for a particle of diameter $125\pm5\,$nm is $31.7\pm1.9\,$kHz. The experimental value is slightly lower, maybe due to imperfect positioning of the nanosphere on the cavity axis. 

\section{One-dimensional quantum ground state in the strong coupling regime}
\label{one-dimensional}

The radiation scattered by the nanosphere on the cavity mode and transmitted by the cavity output mirror is the only sensitive probe that we have to analyze the motion of the nanosphere. Most optomechanical experiments deal with a single mechanical mode weakly coupled to the probe field, a situation that allows to infer some properties of the motion in a direct way. The signature of the mechanical mode is typically a Lorentzian peak in the field spectrum, whose calibrated area yields the displacement variance and the temperature of the mechanical oscillator, while the asymmetry between the Stokes and anti-Stokes peaks in the motional sidebands is a quantum feature allowing a direct measurement of the phononic occupation number \cite{Safavi2012,Khalili2012,Borkje:2016}. In our case the 2D motion, where the corresponding oscillators are coupled through the field, and the strong optomechanical coupling prevent such a direct analysis, which would lead to erroneous results. Even if the (approximated) analytical expressions are obviously useful to understand the system behavior and guide the experiments, a correct quantitative study requires the comparison of the experimental spectra with those generated by a full model, the consequent extraction of the system parameters, and finally the inference of the characteristics of the motion (in particular, the effective thermal occupancy) from the model. The above Section shows that our system is well understood and, in particular, the noise sources are as expected. However, we remark that, for each signal acquisition, we extract the actual decoherence rate (i.e., the amount of noise) from the spectrum. In this procedure, such quantitative estimate of the noise strength critically depends on the reliability of the independent measurement of the detection efficiency. In the classical regime, very similar spectra are obtained by increasing the noise and decreasing the detection efficiency, or vice-versa. The situation changes close to the quantum regime, where the spectral asymmetry between the Stokes and anti-Stokes motional sidebands in the electromagnetic field, that is independent from the detection efficiency, gives a further crucial indication on the achieved quantum state. Even if the direct measurement of the phononic occupation number, typically performed for single-mode systems, is no more possible or reliable, the observation of such spectral asymmetry has the dual purpose of demonstrating the achievement of the quantum regime, and confirming the accuracy of the noise evaluation.  

\begin{figure}
    \centering
    \includegraphics[width=0.75\columnwidth]{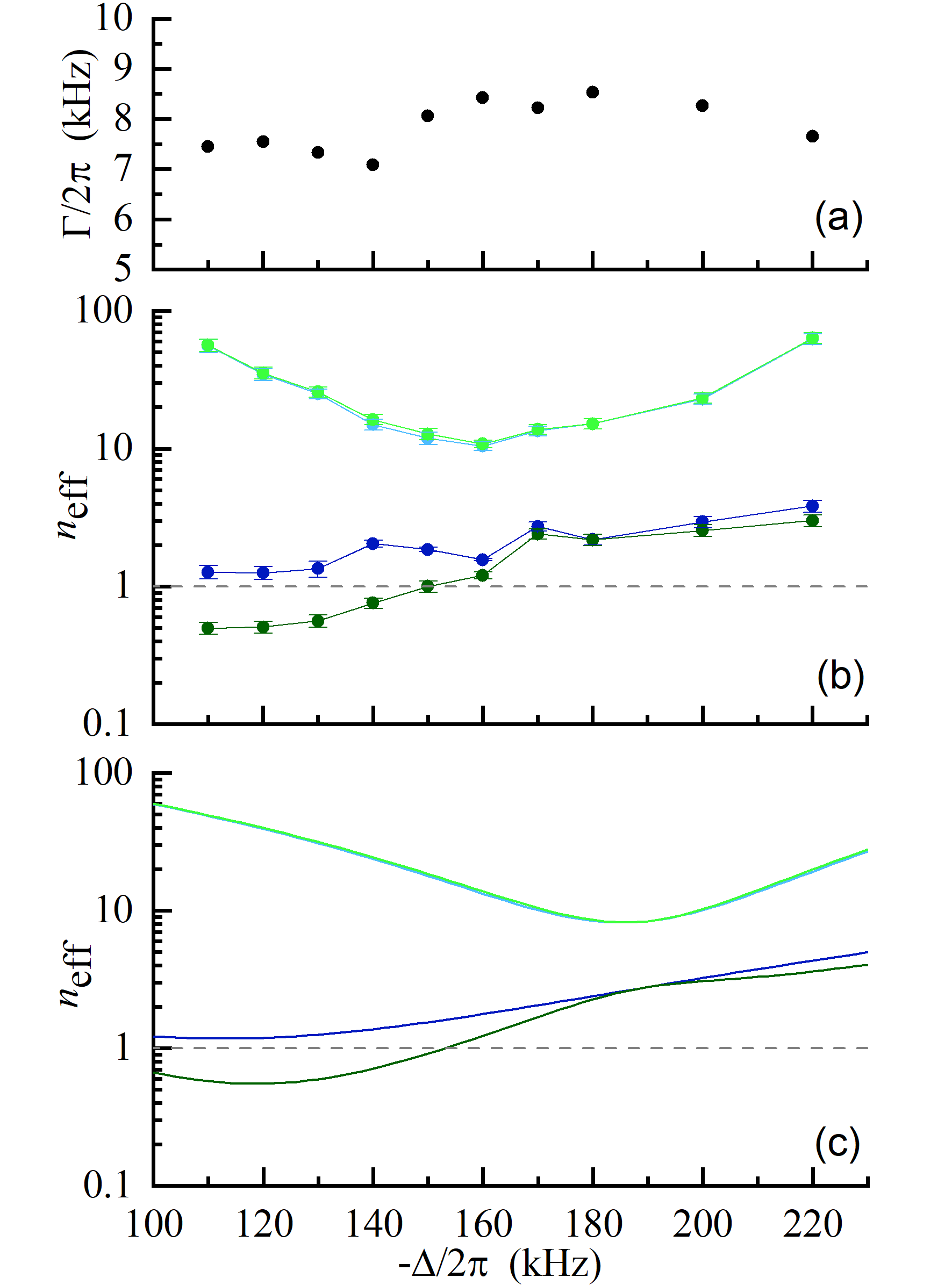}
    \caption{(a) Mean decoherence rate $\, \Gamma =0.5 \,(\GX + \GY)\,$ deduced from the experimental spectra at varying detuning. (b) Effective thermal occupancy along the coldest (dark green), warmest (light green), $X$ (dark blue) and $Y$ (light blue) direction of oscillation, as a function of the detuning of the tweezer light from cavity resonance.
    Solid lines are guides for the eye.  
    (c) Expected trend of the same occupation numbers. These theoretical curves are computed using the following values of the system parameters: $\Omega_X /2 \pi = 125.9 \ \mathrm{kHz}$, $\Omega_Y /2 \pi = 115.95 \ \mathrm{kHz}$, $g_{\mathrm{X}} / 2 \pi = 23.5\, \mathrm{kHz} $, $g_{\mathrm{Y}} / 2 \pi = 3.5\, \mathrm{kHz} $, $\GX /2 \pi = 7.85\, \mathrm{kHz} $, and $\GY /2 \pi = 7.45\, \mathrm{kHz} $ (the decoherence rates are calculated for a background pressure of $P = 7.2 \  10^{-6} \ \mathrm{Pa}$). 
     }
    \label{Fig_nvsdet}
\end{figure}

The measurements at decreasing pressure, described in the previous Section, were performed at a relatively large detuning, yielding moderate optical cooling and stronger signal. When tuning the tweezer radiation closer to resonance, the optomechanical interaction is increased, and the 2D motion is better modeled by a bright mechanical mode and a dark mode. The former corresponds to the motion in the direction of the cavity axis $\XC$ (close to $X$ for small $\theta$), the latter is in the orthogonal direction (i.e., along $\YC$). The decoherence rates measured at different detunings, at low pressure, remain stable as shown in Fig. (\ref{Fig_nvsdet}a), in particular for what concerns their mean value $0.5 (\GX + \GY)$. Indeed, when the two mechanical modes are coupled through the cavity field and their spectra are superposed, we can hardly distinguish the contributions of the two decoherence channels from the shape of the heterodyne spectrum, while the spectral area is determined by their overall effect.

\begin{figure}
    \centering
    \includegraphics{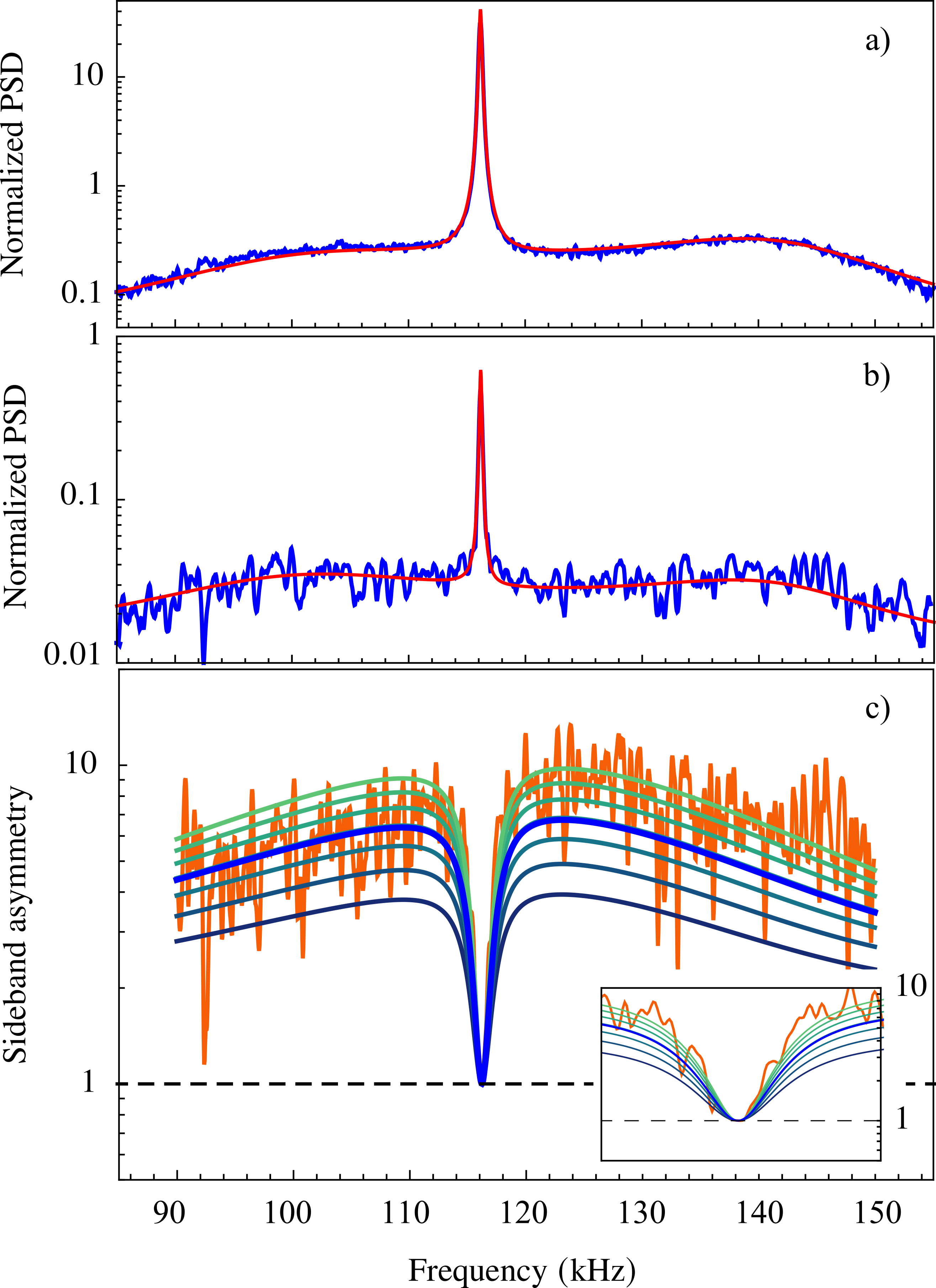}
    \caption{Right (anti-Stokes) (a) and left (Stokes) (b) motional sidebands in the heterodyne spectrum, for a tweezer light detuning of $-\Delta / 2 \pi = 120 \,\mathrm{kHz} $, at a background pressure of $P = 7.2 \times 10^{-5}\,\mathrm{Pa}$. The red solid line is the fit with the theoretical model. 
The inferred values of the optomechanical coupling factors are $\gX/ 2\pi = 24.7 \ \mathrm{kHz}$ and $\gY/ 2 \pi = 4.1 \ \mathrm{kHz}$, hence the system can be described by an oscillator in the strong optomechanical coupling regime, plus a second weakly coupled oscillator. 
    (c) Sideband asymmetry for the heterodyne spectrum, defined in Eq. (\ref{Eq_asym}). The solid lines represent the expected asymmetry for different efficiency values: from $\eta = 0.15$ (lowest curve), up to $\eta = 0.45$ (highest curve), in steps of 0.05. The thickest blue line corresponds to the independently measured efficiency of $\eta = 0.295$. The inset shows an enlarged view of the dip at 116 kHz.
    }
    \label{Fig_asym}
\end{figure} 

For a detuning $\Delta \simeq -\Omega_{\scriptscriptstyle{X,Y}}$ the optomechanical effect is maximum and the 
bright mode is strongly coupled with the optical field, yielding hybridized modes (polaritons) \cite
{quidant2020,Ranfagni2021}. An example of the heterodyne spectrum is shown in Fig. (\ref{Fig_asym}a,b), for 
the anti-Stokes and Stokes sidebands. The two polaritonic peaks are centered at $\sim 100 \,\mathrm{kHz}$ and $\sim 
142 \, \mathrm{kHz}$ (their frequency splitting is $\sim 2 \gX /2\pi$) and are wide (full width) respectively $\sim 
32 \,\mathrm{kHz}$ and $\sim 24.7 \,\mathrm{kHz}$ (i.e., close to $0.5 \kappa/2\pi$). Between the two polaritonic 
peaks, a third narrow peak is visible, centered at $\sim 116 \,\mathrm{kHz}$  and $\sim 120 \,\mathrm{Hz}$ wide. It originates from the bright mode's coupling to the dark mode due to unequal frequencies $\OmegaX \ne \OmegaY$.   

The bright mode is strongly cooled, eventually entering the quantum regime whose signature, as already discussed, is the spectral asymmetry. In the radiation transmitted by the cavity, the motional sidebands are filtered by the cavity transmission function (optical susceptibility). To recover a more meaningful indicator, we define a corrected asymmetry as
\begin{equation}
A(\Omega)\,=\,\frac{S_{\mathrm{out}} (\Omega_{\mathrm{LO}}-\Omega)\,-\,1}{S_{\mathrm{out}} (\Omega_{\mathrm{LO}}+\Omega)\,-\,1}\,\,\frac{(\Omega-\Delta)^2+(\kappa/2)^2}{(\Omega+\Delta)^2+(\kappa/2)^2}  \, ,
\label{Eq_asym}
\end{equation}
which equals the asymmetry in the bright mode spectrum.
This corrected asymmetry is shown in Fig. (\ref{Fig_asym}c). The details of its shape will be discussed in a 
dedicated work, but for the present purposes we underline the following few features. a) A dip at $\sim 116\,$
kHz, very close to the peak frequency of the warm dark mode, falls down to 1, indicating that the correction for 
the cavity filtering is accurate. This dip is a result of destructive interference due to the bright mode's coupling 
to the dark mode, having a much higher quality factor. This can be viewed as a fully mechanical version of optomechanically induced 
transparency. The interference effect is masked by the large thermal noise of the dark mode in the spectra and is 
thus only visible in the asymmetry. b) The maximum asymmetry, occurring at spectral frequency matching $\sim -\Delta
/2\pi = 120\,$kHz, is around 9, denoting a strong non-classical behavior. c) The asymmetry in correspondence of the 
resonance frequencies of the two polaritonic modes is around 6. In the limit of viewing the polaritons as 
independent quantum harmonic oscillators, this would correspond to polariton occupation numbers $1/(A-1) = 0.2$, 
but this information is not sufficient to determine the bright mode's thermal occupancy.

The different theoretical curves shown in Fig. (\ref{Fig_asym}c) with blue-green lines are calculated at increasing values of the detection efficiency $\eta$. A good agreement with the experimental results is observed for $\eta$ between 0.3 and 0.4. The lowest $\chi ^2$ is achieved at $\eta \simeq 0.25$ for a fit of the model to the spectrum including both sidebands, and at $\eta \simeq 0.37$ for the sideband asymmetry. This range supports the independently measured value of $\eta$ that we are using in our analyses.

Without coupling with the cavity field, the 2D motion of the nanosphere in the plane perpendicular to the tweezer axis is well described by two harmonic oscillators along orthogonal directions ($X$ and $Y$) defined by the tweezer polarization. In any other direction of the plane, the motion is given by a weighted sum of the two oscillations, with a spectrum displaying two peaks, and it cannot be simply reproduced by a single harmonic oscillator. Introducing the strong coupling with the cavity field, the system is generally described by three harmonic oscillators associated to the hybridized modes (vectorial polaritons \cite{Ranfagni2021}), defined by the eigenvectors of the drift matrix. Therefore, in this case, in no direction can the motion be simply associated to a single harmonic oscillator. We can characterize the state of the polaritons by their bosonic occupation number, but we cannot straightforwardly assign a phononic occupation number to the motion in any direction. We note in particular that, in the presence of several resonance peaks in the displacement spectrum, the energy equipartition is not preserved, i.e., $\,m^2 \Omega_0^2 \,\langle x^2 \rangle\neq \langle p^2 \rangle$ for any physically meaningful choice of $\Omega_0$ (here, $x$ and $p$ are the position and momentum along the considered direction). 

The motion of the nanosphere can be meaningfully and uniquely characterized by specifying how far it is from a minimum uncertainty state, i.e., by quantifying the parameter $\frac{4}{\hbar^2} \langle x^2 \rangle \langle p^2 \rangle$. For a quantum mechanical oscillator in a thermal state, this parameter coincides with $(2\nm+1)^2$, where $\nm$ is the mean phononic occupation number. It is therefore natural to define an effective thermal occupancy $\neff$ for the motion along the generic $\phi$ direction, as
\begin{equation}
\neff (\phi)\,=\,\frac{1}{2} \left(\frac{2}{\hbar} \sqrt{\langle x_{\phi}^2 \rangle \langle p_{\phi}^2 \rangle} - 1 \right)
\end{equation}
where $\,x_{\phi} = x \sin \phi + y \cos \phi$, $\,p_{\phi} = p_x \sin \phi + p_y \cos \phi$, and $x$, $y$, $p_x$, and $p_y$ are the physical coordinates and momenta along the $X$ and $Y$ direction. These variables are derived from the $\hat{b}_j$ operators used in the model according to the expressions (\ref{eq_xpx}). 
The variances can be calculated as integrals of the spectra, according to $\,\langle \mathcal{O}^2 \rangle = \int S_{\mathcal{O}\mathcal{O}}(\Omega) \frac{\ud\Omega}{2\pi}$.
We remark that our definition of $\neff$ can also be obtained by replacing $\frac{\langle x^2 \rangle}{\xzpf^2}\,\rightarrow \sqrt{\frac{\langle x^2 \rangle}{\xzpf^2}\,\frac{\langle p^2 \rangle}{\pzpf^2}}$ in the usual expression \cite{bowen}: $2 \nm + 1 = \frac{\langle x^2 \rangle}{\xzpf^2} = \frac{1}{\xzpf^2}\int S_{xx} \frac{\ud\Omega}{2\pi}$.

For each set of system parameters, derived from the fit of the output spectrum, we use the theoretical model to calculate the effective thermal occupancy along the $X$ and $Y$ directions (respectively $\neff(\pi/2)$ and $\neff(0)$), as well as its minimum ($\nmin$) and maximum ($\nmax$) values.
With the parameters inferred from the fit of the spectrum shown in Fig. (\ref{Fig_asym}), the minimum effective thermal occupancy is $\nmin = 0.51 \pm 0.05$, where the uncertainty is mainly due to the independent measurement of the detection efficiency. We remark that an occupation number below unity is achieved in the optomechanical strong coupling regime, with resolved polaritonic dynamics.

In Fig, (\ref{Fig_nvsdet}) we show the occupation numbers $\nmin$, $\nmax$, $\neff(0)$ and $\neff(\pi/2)$ for different values of the detuning. Each experimental data point is calculated using the parameters (oscillation frequencies, optomechanical coupling factors and decoherence rates) inferred from the corresponding heterodyne spectrum, while the theoretical curves (panel c) use mean parameters and just vary the detuning, to show the expected behavior.  

\section{Two-dimensional cooling}

\begin{figure}
    \centering
    \includegraphics[width=0.75\columnwidth]{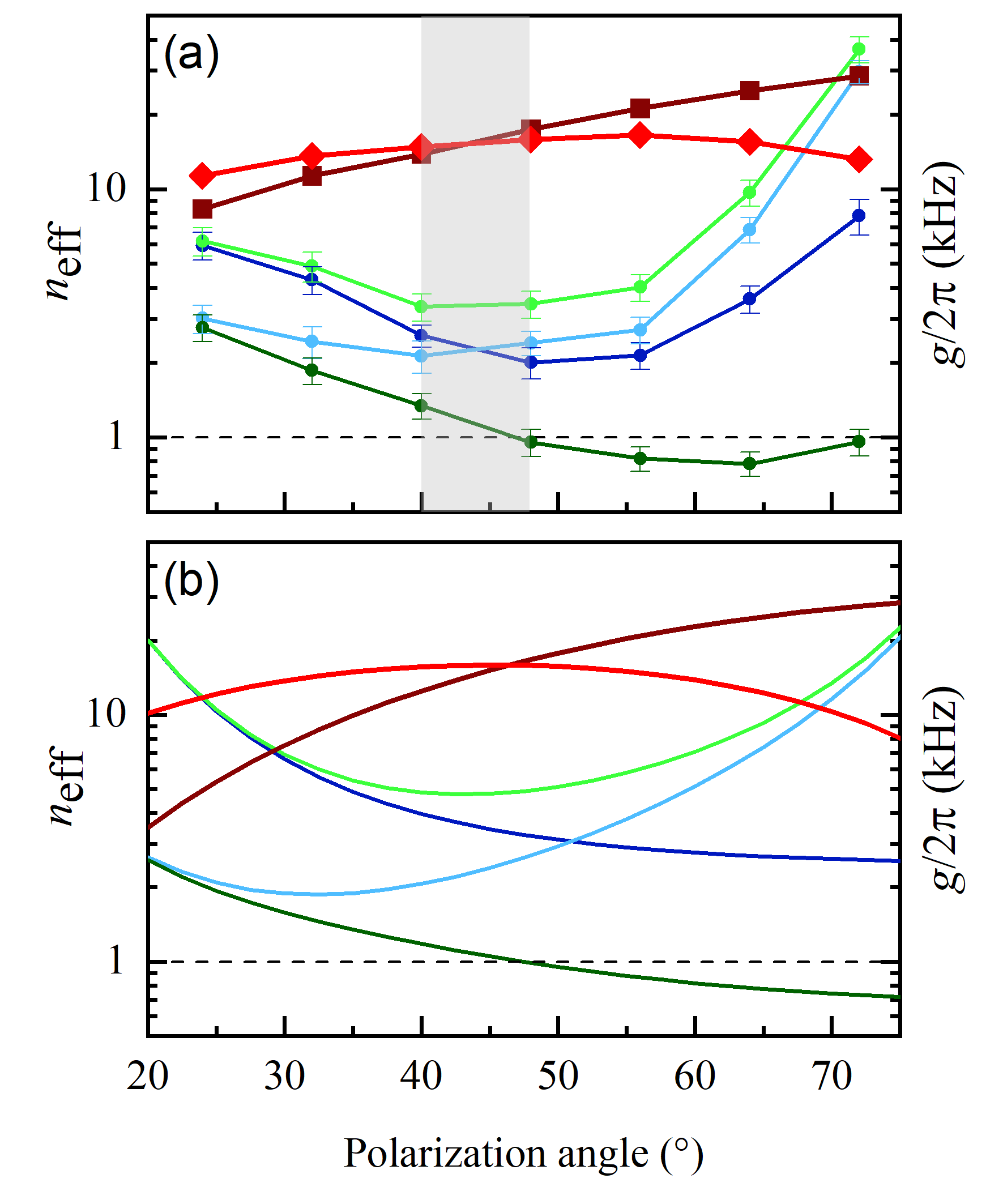}
    \caption{(a) Effective thermal occupancy along the coldest (dark green), warmest (light green), $X$ (dark blue) and $Y$ (light blue) direction of oscillation for a tweezer light detuning of $-\Delta / 2 \pi = 130\, \mathrm{kHz}$, at varying tweezer polarization angle.
    Square, dark red (diamond, light red) symbols: optomechanical coupling coefficient for the $X$ ($Y$) motion. Solid lines are guides for the eye. In the abscissa, we report the polarization angle at the input of the tweezer fiber. The shadowed region highlights the achievement of the strongest 2D confinement.
    (b) Expected trend of the same parameters, plotted as a function of the tweezer polarization angle $\theta$, assuming linear polarization. These theoretical curves are computed using the following values of the system parameters: $-\Delta / 2 \pi =  130 \ \mathrm{kHz}$, $\Omega_X /2 \pi = 125 \ \mathrm{kHz}$, $\Omega_Y /2 \pi = 114.4 \ \mathrm{kHz}$, $g_{\mathrm{max}} / 2 \pi = 31\, \mathrm{kHz} $, $\GX /2 \pi = 12.4\, \mathrm{kHz} $, and $\GY /2 \pi = 12.3\, \mathrm{kHz} $ (the decoherence rates are calculated for a background pressure of $P =1.4 \  10^{-5} \ \mathrm{Pa}$). The agreement with the data of the upper panel is good, yet we remark that the comparison can just be qualitative since the propagation in the tweezer optical fiber modifies the light polarization, yielding an increasing ellipticity and a poorly controlled output angle.
     }
    \label{Fig_nvsangle}
\end{figure}

The spectrum shown in the previous section, as well as those at varying detuning, are acquired for a polarization angle close to $90^{\circ }$, 
such that the direction of strongest cooling (the bright mode direction) is close to the direction $X$ defined by the optical trap. On the other hand, the $Y$ direction is close to the dark mode direction and thus very weakly coupled to the field.
By rotating the polarization angle $\theta$ we increase $\gY$ and enter the regime of true 2D cooling. As shown in Fig. (\ref{Fig_nvsangle}), $\gY$ matches and then overtakes $\gX$, as expected. In the region where the two gains are similar, we also observe the most efficient 2D cooling, i.e., we achieve the lowest value of $\nmax$, which is $3.4\pm 0.4$. At the same time, the smallest effective thermal occupancy $\nmin$ is around 1. This result represents a major improvement with respect to previous experiments, since parametric cooling just allowed to achieve 2D occupation numbers around 100 \cite{Jain2016}, and strong coherent scattering cooling was previously optimized for the single $X$ direction, while the occupation number of the $Y$ mode was just estimated to be below 100 \cite{delic2020}. 

\begin{figure}
    \centering
    \includegraphics{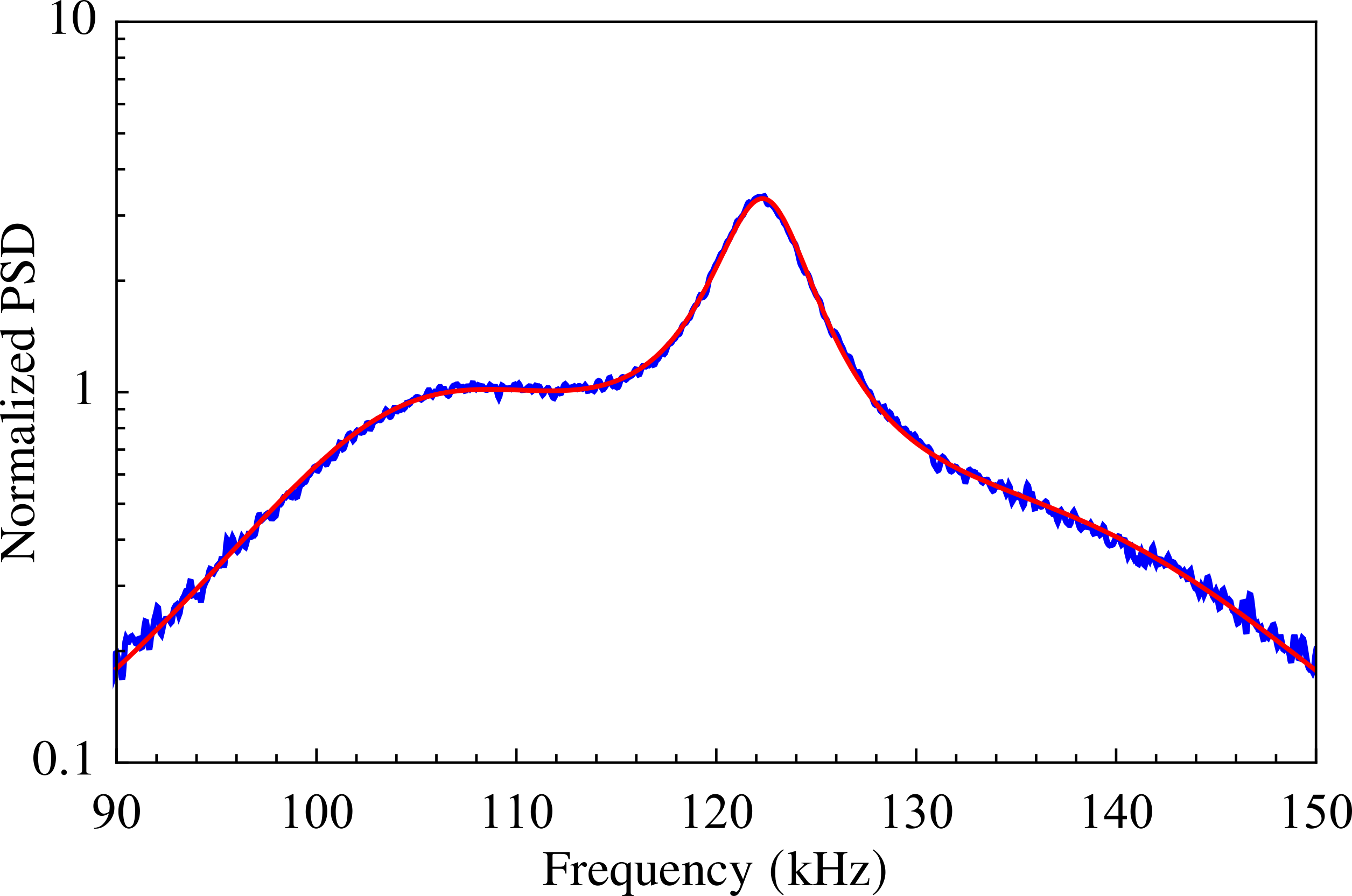}
    \caption{Heterodyne spectrum (anti-Stokes sideband) normalized to the detection shot noise and with the same shot noise subtracted, for the polarization angle that minimizes the effective thermal occupancy in the whole $X-Y$ plane. Red solid line: fit of the theoretical model.
     }
    \label{Fig_sp2}
\end{figure}

In Fig. (\ref{Fig_sp2}) we show the heterodyne spectrum (anti-Stokes sideband) corresponding to the lowest $\nmax$, acquired for a detuning of $\Delta / 2 \pi = -130\, \mathrm{kHz}$, together with the fitted theoretical curve. The optomechanical coupling factors derived from the fit are almost equal, i.e., $\gX / 2 \pi = 13.8\, \mathrm{kHz}$ and $\gY / 2 \pi =  14.8\, \mathrm{kHz}$. We note that the experiment at varying polarization angle was performed with a different nanosphere with respect to the previously described experiences, and $g_{\mathrm{max}}$ is now slightly larger. The system is now in the 2D, coherent strong coupling regime \cite{Ranfagni2021}. The peak originating from the dark mode is broadened, showing that the cooling is indeed efficient in the whole $X-Y$ plane. 

\section{Conclusions}

Coherent scattering in cavity optomechanical experiments is offering a new platform for the creation and coherent control of mesoscopic quantum objects, which recently culminated in the ground-state cooling of a levitated nanosphere along the cavity axis \cite{delic2020}, and in the achievement of quantum-coherent coupling between optical and mechanical modes \cite{Ranfagni2021}.

In this work, we enter the regime in which the fully 2D dynamics of the particle exhibits strong non-classical properties. The motion of the nanosphere is strongly coupled to the cavity field by coherent scattering and cooled in the plane orthogonal to the tweezer axis. 

At very low pressure, the recoil due to photon shot-noise in the scattered radiation play a relevant role in determining the motion of the particle, which at higher pressure is dominated by the effect of collisions with gas molecules. Remarkably, we measure different scattering rates in the $X$ and $Y$ orthogonal spatial directions. Such anisotropy, which is a direct consequence of the dipole radiation pattern, has never been observed so far in mesoscopic objects. 

The accurate measurement of the decoherence rates play a fundamental role in the evaluation of the effective temperature achieved by the particle motion, since standard sideband thermometry is not straightforward.
The heterodyne spectra of the output field exhibit a remarkable asymmetry between the Stokes and anti-Stokes sidebands, revealing the quantum nature of the particle dynamics. Due to the interplay between 2D motion and strong optomechanical coupling, these spectra display a three-peaks structure, ascribable to two polaritonic modes and a dark mode, yielding an asymmetry with a peculiar spectral shape that is not observed in single-mode systems. Hence, we note that it is not possible to infer a mean phononic occupation number from a measured peak asymmetry as customary in optomechanics. This failure of the usual simple characterization procedure is not just a technical problem, but it reveals that the overall physical description requires a novel approach.
Indeed, in the actual 2D landscape, one cannot describe the motion along any direction as a quantum harmonic oscillator, and the strong coupling regime prevents a straightforward definition of a phononic occupation number.
Nevertheless, we show that the motion in any direction can still be characterized in terms of an effective thermal occupancy that quantifies how close the system is to a minimum uncertainty state. This indicator reduces to the usual phonon number for a single quantum oscillator in a thermal state. In our experiment, we achieve a thermal occupancy of $0.51 \pm 0.05$ considering the motion along the coldest direction and, for a suitable polarization angle, a fully 2D cooling with occupancy of $3.4 \pm 0.4$ along the warmest direction and around unity in the orthogonal one.

Our setup thus provides a promising framework for realizing several interesting experiments, with potential applications to quantum information processing and fundamental quantum mechanics.
Of particular interest is the possibility of preparing tripartite optomechanical entangled states involving one optical and two mechanical modes \cite{Hartmann2008,genes2008}. Remarkably, here the two mechanical modes identify the planar motion of the particle, which could enable the preparation of novel quantum states, e.g., involving the coherent superposition of the two motional degrees of freedom, and new forms of directional force and displacement sensing.
We finally remark that the strong 2D cooling represents an important step towards the achievement of three-dimensional motional ground-state cooling which, in combination with quantum-coherent coupling, would allow in the near future the preparation of superposition states of massive objects in free-fall experiments \cite{Romero2011,Bateman2014}, enabling fundamental tests of quantum mechanics, collapse models and quantum gravity  \cite{Gasbarri2021}.

\section*{Acknowledgements}
We thank A. Boschetti for the measurement of the nanoparticles diameter.
Research performed within the Project QuaSeRT (also Research Council of Norway Project Number 285616), funded by the QuantERA ERA-NET
Cofund in Quantum Technologies implemented within the European Union's
Horizon 2020 Programme.

\bibliographystyle{apsrev4-2} 
\bibliography{database}

\begin{thebibliography}{39}%
\makeatletter
\providecommand \@ifxundefined [1]{%
 \@ifx{#1\undefined}
}%
\providecommand \@ifnum [1]{%
 \ifnum #1\expandafter \@firstoftwo
 \else \expandafter \@secondoftwo
 \fi
}%
\providecommand \@ifx [1]{%
 \ifx #1\expandafter \@firstoftwo
 \else \expandafter \@secondoftwo
 \fi
}%
\providecommand \natexlab [1]{#1}%
\providecommand \enquote  [1]{``#1''}%
\providecommand \bibnamefont  [1]{#1}%
\providecommand \bibfnamefont [1]{#1}%
\providecommand \citenamefont [1]{#1}%
\providecommand \href@noop [0]{\@secondoftwo}%
\providecommand \href [0]{\begingroup \@sanitize@url \@href}%
\providecommand \@href[1]{\@@startlink{#1}\@@href}%
\providecommand \@@href[1]{\endgroup#1\@@endlink}%
\providecommand \@sanitize@url [0]{\catcode `\\12\catcode `\$12\catcode
  `\&12\catcode `\#12\catcode `\^12\catcode `\_12\catcode `\%12\relax}%
\providecommand \@@startlink[1]{}%
\providecommand \@@endlink[0]{}%
\providecommand \url  [0]{\begingroup\@sanitize@url \@url }%
\providecommand \@url [1]{\endgroup\@href {#1}{\urlprefix }}%
\providecommand \urlprefix  [0]{URL }%
\providecommand \Eprint [0]{\href }%
\providecommand \doibase [0]{https://doi.org/}%
\providecommand \selectlanguage [0]{\@gobble}%
\providecommand \bibinfo  [0]{\@secondoftwo}%
\providecommand \bibfield  [0]{\@secondoftwo}%
\providecommand \translation [1]{[#1]}%
\providecommand \BibitemOpen [0]{}%
\providecommand \bibitemStop [0]{}%
\providecommand \bibitemNoStop [0]{.\EOS\space}%
\providecommand \EOS [0]{\spacefactor3000\relax}%
\providecommand \BibitemShut  [1]{\csname bibitem#1\endcsname}%
\let\auto@bib@innerbib\@empty
%</preamble>
\bibitem [{\citenamefont {Ashkin}(1970)}]{Ashkin:1970}%
  \BibitemOpen
  \bibfield  {author} {\bibinfo {author} {\bibfnamefont {A.}~\bibnamefont
  {Ashkin}},\ }\href {https://doi.org/10.1103/PhysRevLett.24.156} {\bibfield
  {journal} {\bibinfo  {journal} {Phys. Rev. Lett.}\ }\textbf {\bibinfo
  {volume} {24}},\ \bibinfo {pages} {156} (\bibinfo {year} {1970})}\BibitemShut
  {NoStop}%
\bibitem [{\citenamefont {Chang}\ \emph {et~al.}(2010)\citenamefont {Chang},
  \citenamefont {Regal}, \citenamefont {Papp}, \citenamefont {Wilson},
  \citenamefont {Ye}, \citenamefont {Painter}, \citenamefont {Kimble},\ and\
  \citenamefont {Zoller}}]{Chang:2010}%
  \BibitemOpen
  \bibfield  {author} {\bibinfo {author} {\bibfnamefont {D.~E.}\ \bibnamefont
  {Chang}}, \bibinfo {author} {\bibfnamefont {C.~A.}\ \bibnamefont {Regal}},
  \bibinfo {author} {\bibfnamefont {S.~B.}\ \bibnamefont {Papp}}, \bibinfo
  {author} {\bibfnamefont {D.~J.}\ \bibnamefont {Wilson}}, \bibinfo {author}
  {\bibfnamefont {J.}~\bibnamefont {Ye}}, \bibinfo {author} {\bibfnamefont
  {O.}~\bibnamefont {Painter}}, \bibinfo {author} {\bibfnamefont {H.~J.}\
  \bibnamefont {Kimble}},\ and\ \bibinfo {author} {\bibfnamefont
  {P.}~\bibnamefont {Zoller}},\ }\href
  {https://doi.org/10.1073/pnas.0912969107} {\bibfield  {journal} {\bibinfo
  {journal} {Proceedings of the National Academy of Sciences}\ }\textbf
  {\bibinfo {volume} {107}},\ \bibinfo {pages} {1005} (\bibinfo {year}
  {2010})},\ \Eprint
  {https://arxiv.org/abs/https://www.pnas.org/content/107/3/1005.full.pdf}
  {https://www.pnas.org/content/107/3/1005.full.pdf} \BibitemShut {NoStop}%
\bibitem [{\citenamefont {Barker}\ and\ \citenamefont
  {Shneider}(2010)}]{Barker:2010}%
  \BibitemOpen
  \bibfield  {author} {\bibinfo {author} {\bibfnamefont {P.~F.}\ \bibnamefont
  {Barker}}\ and\ \bibinfo {author} {\bibfnamefont {M.~N.}\ \bibnamefont
  {Shneider}},\ }\href {https://doi.org/10.1103/PhysRevA.81.023826} {\bibfield
  {journal} {\bibinfo  {journal} {Phys. Rev. A}\ }\textbf {\bibinfo {volume}
  {81}},\ \bibinfo {pages} {023826} (\bibinfo {year} {2010})}\BibitemShut
  {NoStop}%
\bibitem [{\citenamefont {Romero-Isart}\ \emph {et~al.}(2010)\citenamefont
  {Romero-Isart}, \citenamefont {Juan}, \citenamefont {Quidant},\ and\
  \citenamefont {Cirac}}]{Romero:2010}%
  \BibitemOpen
  \bibfield  {author} {\bibinfo {author} {\bibfnamefont {O.}~\bibnamefont
  {Romero-Isart}}, \bibinfo {author} {\bibfnamefont {M.}~\bibnamefont {Juan}},
  \bibinfo {author} {\bibfnamefont {R.}~\bibnamefont {Quidant}},\ and\ \bibinfo
  {author} {\bibfnamefont {J.}~\bibnamefont {Cirac}},\ }\href
  {https://doi.org/10.1088/1367-2630/12/3/033015} {\bibfield  {journal}
  {\bibinfo  {journal} {New Journal of Physics}\ }\textbf {\bibinfo {volume}
  {12}},\ \bibinfo {pages} {033015} (\bibinfo {year} {2010})}\BibitemShut
  {NoStop}%
\bibitem [{\citenamefont {Li}\ \emph {et~al.}(2011)\citenamefont {Li},
  \citenamefont {Kheifets},\ and\ \citenamefont {Raizen}}]{Li:2011}%
  \BibitemOpen
  \bibfield  {author} {\bibinfo {author} {\bibfnamefont {T.}~\bibnamefont
  {Li}}, \bibinfo {author} {\bibfnamefont {S.}~\bibnamefont {Kheifets}},\ and\
  \bibinfo {author} {\bibfnamefont {M.}~\bibnamefont {Raizen}},\ }\href
  {https://doi.org/10.1038/nphys1952} {\bibfield  {journal} {\bibinfo
  {journal} {Nature Physics}\ }\textbf {\bibinfo {volume} {7}} (\bibinfo {year}
  {2011})}\BibitemShut {NoStop}%
\bibitem [{\citenamefont {Millen}\ \emph {et~al.}(2020)\citenamefont {Millen},
  \citenamefont {Monteiro}, \citenamefont {Pettit},\ and\ \citenamefont
  {Vamivakas}}]{Millen2020}%
  \BibitemOpen
  \bibfield  {author} {\bibinfo {author} {\bibfnamefont {J.}~\bibnamefont
  {Millen}}, \bibinfo {author} {\bibfnamefont {T.~S.}\ \bibnamefont
  {Monteiro}}, \bibinfo {author} {\bibfnamefont {R.}~\bibnamefont {Pettit}},\
  and\ \bibinfo {author} {\bibfnamefont {A.~N.}\ \bibnamefont {Vamivakas}},\
  }\href {https://doi.org/10.1088/1361-6633/ab6100} {\bibfield  {journal}
  {\bibinfo  {journal} {Reports on Progress in Physics}\ }\textbf {\bibinfo
  {volume} {83}},\ \bibinfo {pages} {026401} (\bibinfo {year}
  {2020})}\BibitemShut {NoStop}%
\bibitem [{\citenamefont {Magrini}\ \emph {et~al.}(2021)\citenamefont
  {Magrini}, \citenamefont {Rosenzweig}, \citenamefont {Bach}, \citenamefont
  {Deutschmann-Olek}, \citenamefont {Hofer}, \citenamefont {Hong},
  \citenamefont {Kiesel}, \citenamefont {Kugi},\ and\ \citenamefont
  {Aspelmeyer}}]{Magrini2021}%
  \BibitemOpen
  \bibfield  {author} {\bibinfo {author} {\bibfnamefont {L.}~\bibnamefont
  {Magrini}}, \bibinfo {author} {\bibfnamefont {P.}~\bibnamefont {Rosenzweig}},
  \bibinfo {author} {\bibfnamefont {C.}~\bibnamefont {Bach}}, \bibinfo {author}
  {\bibfnamefont {A.}~\bibnamefont {Deutschmann-Olek}}, \bibinfo {author}
  {\bibfnamefont {S.~G.}\ \bibnamefont {Hofer}}, \bibinfo {author}
  {\bibfnamefont {S.}~\bibnamefont {Hong}}, \bibinfo {author} {\bibfnamefont
  {N.}~\bibnamefont {Kiesel}}, \bibinfo {author} {\bibfnamefont
  {A.}~\bibnamefont {Kugi}},\ and\ \bibinfo {author} {\bibfnamefont
  {M.}~\bibnamefont {Aspelmeyer}},\ }\href
  {https://doi.org/10.1038/s41586-021-03602-3} {\bibfield  {journal} {\bibinfo
  {journal} {Nature}\ }\textbf {\bibinfo {volume} {595}},\ \bibinfo {pages}
  {373} (\bibinfo {year} {2021})}\BibitemShut {NoStop}%
\bibitem [{\citenamefont {Tebbenjohanns}\ \emph {et~al.}(2021)\citenamefont
  {Tebbenjohanns}, \citenamefont {Mattana}, \citenamefont {Rossi},
  \citenamefont {Frimmer},\ and\ \citenamefont {Novotny}}]{Tebbenjohanns2021}%
  \BibitemOpen
  \bibfield  {author} {\bibinfo {author} {\bibfnamefont {F.}~\bibnamefont
  {Tebbenjohanns}}, \bibinfo {author} {\bibfnamefont {M.~L.}\ \bibnamefont
  {Mattana}}, \bibinfo {author} {\bibfnamefont {M.}~\bibnamefont {Rossi}},
  \bibinfo {author} {\bibfnamefont {M.}~\bibnamefont {Frimmer}},\ and\ \bibinfo
  {author} {\bibfnamefont {L.}~\bibnamefont {Novotny}},\ }\href
  {https://doi.org/10.1038/s41586-021-03617-w} {\bibfield  {journal} {\bibinfo
  {journal} {Nature}\ }\textbf {\bibinfo {volume} {595}},\ \bibinfo {pages}
  {378} (\bibinfo {year} {2021})}\BibitemShut {NoStop}%
\bibitem [{\citenamefont {Deli{\'c}}\ \emph {et~al.}(2020)\citenamefont
  {Deli{\'c}}, \citenamefont {Reisenbauer}, \citenamefont {Dare}, \citenamefont
  {Grass}, \citenamefont {Vuleti{\'c}}, \citenamefont {Kiesel},\ and\
  \citenamefont {Aspelmeyer}}]{delic2020}%
  \BibitemOpen
  \bibfield  {author} {\bibinfo {author} {\bibfnamefont {U.}~\bibnamefont
  {Deli{\'c}}}, \bibinfo {author} {\bibfnamefont {M.}~\bibnamefont
  {Reisenbauer}}, \bibinfo {author} {\bibfnamefont {K.}~\bibnamefont {Dare}},
  \bibinfo {author} {\bibfnamefont {D.}~\bibnamefont {Grass}}, \bibinfo
  {author} {\bibfnamefont {V.}~\bibnamefont {Vuleti{\'c}}}, \bibinfo {author}
  {\bibfnamefont {N.}~\bibnamefont {Kiesel}},\ and\ \bibinfo {author}
  {\bibfnamefont {M.}~\bibnamefont {Aspelmeyer}},\ }\href
  {https://doi.org/10.1126/science.aba3993} {\bibfield  {journal} {\bibinfo
  {journal} {Science}\ }\textbf {\bibinfo {volume} {367}},\ \bibinfo {pages}
  {892} (\bibinfo {year} {2020})},\ \Eprint
  {https://arxiv.org/abs/https://science.sciencemag.org/content/367/6480/892.full.pdf}
  {https://science.sciencemag.org/content/367/6480/892.full.pdf} \BibitemShut
  {NoStop}%
\bibitem [{\citenamefont {Kiesel}\ \emph {et~al.}(2013)\citenamefont {Kiesel},
  \citenamefont {Blaser}, \citenamefont {Deli{\'c}}, \citenamefont {Grass},
  \citenamefont {Kaltenbaek},\ and\ \citenamefont {Aspelmeyer}}]{Kiesel:2013}%
  \BibitemOpen
  \bibfield  {author} {\bibinfo {author} {\bibfnamefont {N.}~\bibnamefont
  {Kiesel}}, \bibinfo {author} {\bibfnamefont {F.}~\bibnamefont {Blaser}},
  \bibinfo {author} {\bibfnamefont {U.}~\bibnamefont {Deli{\'c}}}, \bibinfo
  {author} {\bibfnamefont {D.}~\bibnamefont {Grass}}, \bibinfo {author}
  {\bibfnamefont {R.}~\bibnamefont {Kaltenbaek}},\ and\ \bibinfo {author}
  {\bibfnamefont {M.}~\bibnamefont {Aspelmeyer}},\ }\href
  {https://doi.org/10.1073/pnas.1309167110} {\bibfield  {journal} {\bibinfo
  {journal} {Proceedings of the National Academy of Sciences}\ }\textbf
  {\bibinfo {volume} {110}},\ \bibinfo {pages} {14180} (\bibinfo {year}
  {2013})},\ \Eprint
  {https://arxiv.org/abs/https://www.pnas.org/content/110/35/14180.full.pdf}
  {https://www.pnas.org/content/110/35/14180.full.pdf} \BibitemShut {NoStop}%
\bibitem [{\citenamefont {Deli\'{c}}\ \emph {et~al.}(2019)\citenamefont
  {Deli\'{c}}, \citenamefont {Reisenbauer}, \citenamefont {Grass},
  \citenamefont {Kiesel}, \citenamefont {Vuletic},\ and\ \citenamefont
  {Aspelmeyer}}]{delic2019}%
  \BibitemOpen
  \bibfield  {author} {\bibinfo {author} {\bibfnamefont {U.}~\bibnamefont
  {Deli\'{c}}}, \bibinfo {author} {\bibfnamefont {M.}~\bibnamefont
  {Reisenbauer}}, \bibinfo {author} {\bibfnamefont {D.}~\bibnamefont {Grass}},
  \bibinfo {author} {\bibfnamefont {N.}~\bibnamefont {Kiesel}}, \bibinfo
  {author} {\bibfnamefont {V.}~\bibnamefont {Vuletic}},\ and\ \bibinfo {author}
  {\bibfnamefont {M.}~\bibnamefont {Aspelmeyer}},\ }\href
  {https://doi.org/10.1103/PhysRevLett.122.123602} {\bibfield  {journal}
  {\bibinfo  {journal} {Physical Review Letters}\ }\textbf {\bibinfo {volume}
  {122}},\ \bibinfo {pages} {123602} (\bibinfo {year} {2019})}\BibitemShut
  {NoStop}%
\bibitem [{\citenamefont {Windey}\ \emph {et~al.}(2019)\citenamefont {Windey},
  \citenamefont {Gonzalez-Ballestero}, \citenamefont {Maurer}, \citenamefont
  {Novotny}, \citenamefont {Romero-Isart},\ and\ \citenamefont
  {Reimann}}]{windey2019}%
  \BibitemOpen
  \bibfield  {author} {\bibinfo {author} {\bibfnamefont {D.}~\bibnamefont
  {Windey}}, \bibinfo {author} {\bibfnamefont {C.}~\bibnamefont
  {Gonzalez-Ballestero}}, \bibinfo {author} {\bibfnamefont {P.}~\bibnamefont
  {Maurer}}, \bibinfo {author} {\bibfnamefont {L.}~\bibnamefont {Novotny}},
  \bibinfo {author} {\bibfnamefont {O.}~\bibnamefont {Romero-Isart}},\ and\
  \bibinfo {author} {\bibfnamefont {R.}~\bibnamefont {Reimann}},\ }\href
  {https://doi.org/10.1103/PhysRevLett.122.123601} {\bibfield  {journal}
  {\bibinfo  {journal} {Phys. Rev. Lett.}\ }\textbf {\bibinfo {volume} {122}},\
  \bibinfo {pages} {123601} (\bibinfo {year} {2019})}\BibitemShut {NoStop}%
\bibitem [{\citenamefont {Gonzalez-Ballestero}\ \emph
  {et~al.}(2019)\citenamefont {Gonzalez-Ballestero}, \citenamefont {Maurer},
  \citenamefont {Windey}, \citenamefont {Novotny}, \citenamefont {Reimann},\
  and\ \citenamefont {Romero-Isart}}]{gonzalez-ballestrero2019}%
  \BibitemOpen
  \bibfield  {author} {\bibinfo {author} {\bibfnamefont {C.}~\bibnamefont
  {Gonzalez-Ballestero}}, \bibinfo {author} {\bibfnamefont {P.}~\bibnamefont
  {Maurer}}, \bibinfo {author} {\bibfnamefont {D.}~\bibnamefont {Windey}},
  \bibinfo {author} {\bibfnamefont {L.}~\bibnamefont {Novotny}}, \bibinfo
  {author} {\bibfnamefont {R.}~\bibnamefont {Reimann}},\ and\ \bibinfo {author}
  {\bibfnamefont {O.}~\bibnamefont {Romero-Isart}},\ }\href
  {https://doi.org/10.1103/PhysRevA.100.013805} {\bibfield  {journal} {\bibinfo
   {journal} {Phys. Rev. A}\ }\textbf {\bibinfo {volume} {100}},\ \bibinfo
  {pages} {013805} (\bibinfo {year} {2019})}\BibitemShut {NoStop}%
\bibitem [{\citenamefont {Vuleti\ifmmode~\acute{c}\else \'{c}\fi{}}\ and\
  \citenamefont {Chu}(2000)}]{Vuletic2000}%
  \BibitemOpen
  \bibfield  {author} {\bibinfo {author} {\bibfnamefont {V.}~\bibnamefont
  {Vuleti\ifmmode~\acute{c}\else \'{c}\fi{}}}\ and\ \bibinfo {author}
  {\bibfnamefont {S.}~\bibnamefont {Chu}},\ }\href
  {https://doi.org/10.1103/PhysRevLett.84.3787} {\bibfield  {journal} {\bibinfo
   {journal} {Phys. Rev. Lett.}\ }\textbf {\bibinfo {volume} {84}},\ \bibinfo
  {pages} {3787} (\bibinfo {year} {2000})}\BibitemShut {NoStop}%
\bibitem [{\citenamefont {Novotny}\ and\ \citenamefont
  {Hecht}(2012)}]{Novotny:2012}%
  \BibitemOpen
  \bibfield  {author} {\bibinfo {author} {\bibfnamefont {L.}~\bibnamefont
  {Novotny}}\ and\ \bibinfo {author} {\bibfnamefont {B.}~\bibnamefont
  {Hecht}},\ }\href {https://doi.org/10.1017/CBO9780511794193} {\emph {\bibinfo
  {title} {Principles of Nano-Optics}}},\ \bibinfo {edition} {2nd}\ ed.\
  (\bibinfo  {publisher} {Cambridge University Press},\ \bibinfo {year}
  {2012})\BibitemShut {NoStop}%
\bibitem [{\citenamefont {Genes}\ \emph {et~al.}(2008)\citenamefont {Genes},
  \citenamefont {Vitali},\ and\ \citenamefont {Tombesi}}]{genes2008}%
  \BibitemOpen
  \bibfield  {author} {\bibinfo {author} {\bibfnamefont {C.}~\bibnamefont
  {Genes}}, \bibinfo {author} {\bibfnamefont {D.}~\bibnamefont {Vitali}},\ and\
  \bibinfo {author} {\bibfnamefont {P.}~\bibnamefont {Tombesi}},\ }\href
  {https://doi.org/10.1088/1367-2630/10/9/095009} {\bibfield  {journal}
  {\bibinfo  {journal} {New Journal of Physics}\ }\textbf {\bibinfo {volume}
  {10}},\ \bibinfo {pages} {095009} (\bibinfo {year} {2008})}\BibitemShut
  {NoStop}%
\bibitem [{\citenamefont {Massel}\ \emph {et~al.}(2012)\citenamefont {Massel},
  \citenamefont {Cho}, \citenamefont {Pirkkalainen}, \citenamefont {Hakonen},
  \citenamefont {Heikkil{\"a}},\ and\ \citenamefont
  {Sillanp{\"a}{\"a}}}]{massel2012}%
  \BibitemOpen
  \bibfield  {author} {\bibinfo {author} {\bibfnamefont {F.}~\bibnamefont
  {Massel}}, \bibinfo {author} {\bibfnamefont {S.~U.}\ \bibnamefont {Cho}},
  \bibinfo {author} {\bibfnamefont {J.-M.}\ \bibnamefont {Pirkkalainen}},
  \bibinfo {author} {\bibfnamefont {P.~J.}\ \bibnamefont {Hakonen}}, \bibinfo
  {author} {\bibfnamefont {T.~T.}\ \bibnamefont {Heikkil{\"a}}},\ and\ \bibinfo
  {author} {\bibfnamefont {M.~A.}\ \bibnamefont {Sillanp{\"a}{\"a}}},\ }\href
  {https://doi.org/10.1038/ncomms1993} {\bibfield  {journal} {\bibinfo
  {journal} {Nature Communications}\ }\textbf {\bibinfo {volume} {3}},\
  \bibinfo {pages} {987} (\bibinfo {year} {2012})}\BibitemShut {NoStop}%
\bibitem [{\citenamefont {Shkarin}\ \emph {et~al.}(2014)\citenamefont
  {Shkarin}, \citenamefont {Flowers-Jacobs}, \citenamefont {Hoch},
  \citenamefont {Kashkanova}, \citenamefont {Deutsch}, \citenamefont
  {Reichel},\ and\ \citenamefont {Harris}}]{shkarin}%
  \BibitemOpen
  \bibfield  {author} {\bibinfo {author} {\bibfnamefont {A.~B.}\ \bibnamefont
  {Shkarin}}, \bibinfo {author} {\bibfnamefont {N.~E.}\ \bibnamefont
  {Flowers-Jacobs}}, \bibinfo {author} {\bibfnamefont {S.~W.}\ \bibnamefont
  {Hoch}}, \bibinfo {author} {\bibfnamefont {A.~D.}\ \bibnamefont
  {Kashkanova}}, \bibinfo {author} {\bibfnamefont {C.}~\bibnamefont {Deutsch}},
  \bibinfo {author} {\bibfnamefont {J.}~\bibnamefont {Reichel}},\ and\ \bibinfo
  {author} {\bibfnamefont {J.~G.~E.}\ \bibnamefont {Harris}},\ }\href
  {https://doi.org/10.1103/PhysRevLett.112.013602} {\bibfield  {journal}
  {\bibinfo  {journal} {Phys. Rev. Lett.}\ }\textbf {\bibinfo {volume} {112}},\
  \bibinfo {pages} {013602} (\bibinfo {year} {2014})}\BibitemShut {NoStop}%
\bibitem [{\citenamefont {Toro\ifmmode~\check{s}\else \v{s}\fi{}}\ \emph
  {et~al.}(2021)\citenamefont {Toro\ifmmode~\check{s}\else \v{s}\fi{}},
  \citenamefont {Deli\ifmmode~\acute{c}\else \'{c}\fi{}}, \citenamefont
  {Hales},\ and\ \citenamefont {Monteiro}}]{toros2021}%
  \BibitemOpen
  \bibfield  {author} {\bibinfo {author} {\bibfnamefont {M.}~\bibnamefont
  {Toro\ifmmode~\check{s}\else \v{s}\fi{}}}, \bibinfo {author} {\bibfnamefont
  {U.~c.~v.}\ \bibnamefont {Deli\ifmmode~\acute{c}\else \'{c}\fi{}}}, \bibinfo
  {author} {\bibfnamefont {F.}~\bibnamefont {Hales}},\ and\ \bibinfo {author}
  {\bibfnamefont {T.~S.}\ \bibnamefont {Monteiro}},\ }\href
  {https://doi.org/10.1103/PhysRevResearch.3.023071} {\bibfield  {journal}
  {\bibinfo  {journal} {Phys. Rev. Research}\ }\textbf {\bibinfo {volume}
  {3}},\ \bibinfo {pages} {023071} (\bibinfo {year} {2021})}\BibitemShut
  {NoStop}%
\bibitem [{\citenamefont {Ranfagni}\ \emph {et~al.}(2021)\citenamefont
  {Ranfagni}, \citenamefont {Vezio}, \citenamefont {Calamai}, \citenamefont
  {Chowdhury}, \citenamefont {Marino},\ and\ \citenamefont
  {Marin}}]{Ranfagni2021}%
  \BibitemOpen
  \bibfield  {author} {\bibinfo {author} {\bibfnamefont {A.}~\bibnamefont
  {Ranfagni}}, \bibinfo {author} {\bibfnamefont {P.}~\bibnamefont {Vezio}},
  \bibinfo {author} {\bibfnamefont {M.}~\bibnamefont {Calamai}}, \bibinfo
  {author} {\bibfnamefont {A.}~\bibnamefont {Chowdhury}}, \bibinfo {author}
  {\bibfnamefont {F.}~\bibnamefont {Marino}},\ and\ \bibinfo {author}
  {\bibfnamefont {F.}~\bibnamefont {Marin}},\ }\href
  {https://doi.org/10.1038/s41567-021-01307-y} {\bibfield  {journal} {\bibinfo
  {journal} {Nature Physics}\ }\textbf {\bibinfo {volume} {17}},\ \bibinfo
  {pages} {1120} (\bibinfo {year} {2021})}\BibitemShut {NoStop}%
\bibitem [{\citenamefont {Safavi-Naeini}\ \emph {et~al.}(2012)\citenamefont
  {Safavi-Naeini}, \citenamefont {Chan}, \citenamefont {Hill}, \citenamefont
  {Alegre}, \citenamefont {Krause},\ and\ \citenamefont
  {Painter}}]{Safavi2012}%
  \BibitemOpen
  \bibfield  {author} {\bibinfo {author} {\bibfnamefont {A.~H.}\ \bibnamefont
  {Safavi-Naeini}}, \bibinfo {author} {\bibfnamefont {J.}~\bibnamefont {Chan}},
  \bibinfo {author} {\bibfnamefont {J.~T.}\ \bibnamefont {Hill}}, \bibinfo
  {author} {\bibfnamefont {T.~P.~M.}\ \bibnamefont {Alegre}}, \bibinfo {author}
  {\bibfnamefont {A.}~\bibnamefont {Krause}},\ and\ \bibinfo {author}
  {\bibfnamefont {O.}~\bibnamefont {Painter}},\ }\href
  {https://doi.org/10.1103/PhysRevLett.108.033602} {\bibfield  {journal}
  {\bibinfo  {journal} {Phys. Rev. Lett.}\ }\textbf {\bibinfo {volume} {108}},\
  \bibinfo {pages} {033602} (\bibinfo {year} {2012})}\BibitemShut {NoStop}%
\bibitem [{\citenamefont {Khalili}\ \emph {et~al.}(2012)\citenamefont
  {Khalili}, \citenamefont {Miao}, \citenamefont {Yang}, \citenamefont
  {Safavi-Naeini}, \citenamefont {Painter},\ and\ \citenamefont
  {Chen}}]{Khalili2012}%
  \BibitemOpen
  \bibfield  {author} {\bibinfo {author} {\bibfnamefont {F.~Y.}\ \bibnamefont
  {Khalili}}, \bibinfo {author} {\bibfnamefont {H.}~\bibnamefont {Miao}},
  \bibinfo {author} {\bibfnamefont {H.}~\bibnamefont {Yang}}, \bibinfo {author}
  {\bibfnamefont {A.~H.}\ \bibnamefont {Safavi-Naeini}}, \bibinfo {author}
  {\bibfnamefont {O.}~\bibnamefont {Painter}},\ and\ \bibinfo {author}
  {\bibfnamefont {Y.}~\bibnamefont {Chen}},\ }\href
  {https://doi.org/10.1103/PhysRevA.86.033840} {\bibfield  {journal} {\bibinfo
  {journal} {Phys. Rev. A}\ }\textbf {\bibinfo {volume} {86}},\ \bibinfo
  {pages} {033840} (\bibinfo {year} {2012})}\BibitemShut {NoStop}%
\bibitem [{\citenamefont {Borkje}(2016)}]{Borkje:2016}%
  \BibitemOpen
  \bibfield  {author} {\bibinfo {author} {\bibfnamefont {K.}~\bibnamefont
  {Borkje}},\ }\href {https://doi.org/10.1103/PhysRevA.94.043816} {\bibfield
  {journal} {\bibinfo  {journal} {Physical Review A}\ }\textbf {\bibinfo
  {volume} {94}} (\bibinfo {year} {2016})}\BibitemShut {NoStop}%
\bibitem [{\citenamefont {Toro\ifmmode~\check{s}\else \v{s}\fi{}}\ and\
  \citenamefont {Monteiro}(2020)}]{toros2020}%
  \BibitemOpen
  \bibfield  {author} {\bibinfo {author} {\bibfnamefont {M.}~\bibnamefont
  {Toro\ifmmode~\check{s}\else \v{s}\fi{}}}\ and\ \bibinfo {author}
  {\bibfnamefont {T.~S.}\ \bibnamefont {Monteiro}},\ }\href
  {https://doi.org/10.1103/PhysRevResearch.2.023228} {\bibfield  {journal}
  {\bibinfo  {journal} {Phys. Rev. Research}\ }\textbf {\bibinfo {volume}
  {2}},\ \bibinfo {pages} {023228} (\bibinfo {year} {2020})}\BibitemShut
  {NoStop}%
\bibitem [{\citenamefont {Duri}\ \emph {et~al.}(2009)\citenamefont {Duri},
  \citenamefont {Sessoms}, \citenamefont {Trappe},\ and\ \citenamefont
  {Cipelletti}}]{Duri2009}%
  \BibitemOpen
  \bibfield  {author} {\bibinfo {author} {\bibfnamefont {A.}~\bibnamefont
  {Duri}}, \bibinfo {author} {\bibfnamefont {D.~A.}\ \bibnamefont {Sessoms}},
  \bibinfo {author} {\bibfnamefont {V.}~\bibnamefont {Trappe}},\ and\ \bibinfo
  {author} {\bibfnamefont {L.}~\bibnamefont {Cipelletti}},\ }\href
  {https://doi.org/10.1103/PhysRevLett.102.085702} {\bibfield  {journal}
  {\bibinfo  {journal} {Phys. Rev. Lett.}\ }\textbf {\bibinfo {volume} {102}},\
  \bibinfo {pages} {085702} (\bibinfo {year} {2009})}\BibitemShut {NoStop}%
\bibitem [{\citenamefont {Mestres}\ \emph {et~al.}(2015)\citenamefont
  {Mestres}, \citenamefont {Berthelot}, \citenamefont {Spasenović},
  \citenamefont {Gieseler}, \citenamefont {Novotny},\ and\ \citenamefont
  {Quidant}}]{mestres2015}%
  \BibitemOpen
  \bibfield  {author} {\bibinfo {author} {\bibfnamefont {P.}~\bibnamefont
  {Mestres}}, \bibinfo {author} {\bibfnamefont {J.}~\bibnamefont {Berthelot}},
  \bibinfo {author} {\bibfnamefont {M.}~\bibnamefont {Spasenović}}, \bibinfo
  {author} {\bibfnamefont {J.}~\bibnamefont {Gieseler}}, \bibinfo {author}
  {\bibfnamefont {L.}~\bibnamefont {Novotny}},\ and\ \bibinfo {author}
  {\bibfnamefont {R.}~\bibnamefont {Quidant}},\ }\href
  {https://doi.org/10.1063/1.4933180} {\bibfield  {journal} {\bibinfo
  {journal} {Applied Physics Letters}\ }\textbf {\bibinfo {volume} {107}},\
  \bibinfo {pages} {151102} (\bibinfo {year} {2015})},\ \Eprint
  {https://arxiv.org/abs/https://doi.org/10.1063/1.4933180}
  {https://doi.org/10.1063/1.4933180} \BibitemShut {NoStop}%
\bibitem [{\citenamefont {Calamai}\ \emph {et~al.}(2021)\citenamefont
  {Calamai}, \citenamefont {Ranfagni},\ and\ \citenamefont
  {Marin}}]{calamai2020}%
  \BibitemOpen
  \bibfield  {author} {\bibinfo {author} {\bibfnamefont {M.}~\bibnamefont
  {Calamai}}, \bibinfo {author} {\bibfnamefont {A.}~\bibnamefont {Ranfagni}},\
  and\ \bibinfo {author} {\bibfnamefont {F.}~\bibnamefont {Marin}},\ }\href
  {https://doi.org/10.1063/5.0024432} {\bibfield  {journal} {\bibinfo
  {journal} {AIP Advances}\ }\textbf {\bibinfo {volume} {11}},\ \bibinfo
  {pages} {025246} (\bibinfo {year} {2021})},\ \Eprint
  {https://arxiv.org/abs/https://doi.org/10.1063/5.0024432}
  {https://doi.org/10.1063/5.0024432} \BibitemShut {NoStop}%
\bibitem [{\citenamefont {Epstein}(1924)}]{Epstein}%
  \BibitemOpen
  \bibfield  {author} {\bibinfo {author} {\bibfnamefont {P.~S.}\ \bibnamefont
  {Epstein}},\ }\href {https://doi.org/10.1103/PhysRev.23.710} {\bibfield
  {journal} {\bibinfo  {journal} {Phys. Rev.}\ }\textbf {\bibinfo {volume}
  {23}},\ \bibinfo {pages} {710} (\bibinfo {year} {1924})}\BibitemShut
  {NoStop}%
\bibitem [{\citenamefont {Beresnev}\ \emph {et~al.}(1990)\citenamefont
  {Beresnev}, \citenamefont {Chernyak},\ and\ \citenamefont
  {Fomyagin}}]{Beresnev:1990}%
  \BibitemOpen
  \bibfield  {author} {\bibinfo {author} {\bibfnamefont {S.}~\bibnamefont
  {Beresnev}}, \bibinfo {author} {\bibfnamefont {V.}~\bibnamefont {Chernyak}},\
  and\ \bibinfo {author} {\bibfnamefont {G.}~\bibnamefont {Fomyagin}},\ }\href
  {https://doi.org/10.1017/S0022112090003007} {\bibfield  {journal} {\bibinfo
  {journal} {Journal of Fluid Mechanics}\ }\textbf {\bibinfo {volume} {219}},\
  \bibinfo {pages} {405 } (\bibinfo {year} {1990})}\BibitemShut {NoStop}%
\bibitem [{\citenamefont {Seberson}\ and\ \citenamefont
  {Robicheaux}(2020)}]{Seberson2020}%
  \BibitemOpen
  \bibfield  {author} {\bibinfo {author} {\bibfnamefont {T.}~\bibnamefont
  {Seberson}}\ and\ \bibinfo {author} {\bibfnamefont {F.}~\bibnamefont
  {Robicheaux}},\ }\href {https://doi.org/10.1103/PhysRevA.102.033505}
  {\bibfield  {journal} {\bibinfo  {journal} {Phys. Rev. A}\ }\textbf {\bibinfo
  {volume} {102}},\ \bibinfo {pages} {033505} (\bibinfo {year}
  {2020})}\BibitemShut {NoStop}%
\bibitem [{\citenamefont {Jain}\ \emph {et~al.}(2016)\citenamefont {Jain},
  \citenamefont {Gieseler}, \citenamefont {Moritz}, \citenamefont {Dellago},
  \citenamefont {Quidant},\ and\ \citenamefont {Novotny}}]{Jain2016}%
  \BibitemOpen
  \bibfield  {author} {\bibinfo {author} {\bibfnamefont {V.}~\bibnamefont
  {Jain}}, \bibinfo {author} {\bibfnamefont {J.}~\bibnamefont {Gieseler}},
  \bibinfo {author} {\bibfnamefont {C.}~\bibnamefont {Moritz}}, \bibinfo
  {author} {\bibfnamefont {C.}~\bibnamefont {Dellago}}, \bibinfo {author}
  {\bibfnamefont {R.}~\bibnamefont {Quidant}},\ and\ \bibinfo {author}
  {\bibfnamefont {L.}~\bibnamefont {Novotny}},\ }\href
  {https://doi.org/10.1103/PhysRevLett.116.243601} {\bibfield  {journal}
  {\bibinfo  {journal} {Phys. Rev. Lett.}\ }\textbf {\bibinfo {volume} {116}},\
  \bibinfo {pages} {243601} (\bibinfo {year} {2016})}\BibitemShut {NoStop}%
\bibitem [{\citenamefont {Millen}\ \emph {et~al.}(2014)\citenamefont {Millen},
  \citenamefont {Deesuwan}, \citenamefont {Barker},\ and\ \citenamefont
  {Anders}}]{Millen2014}%
  \BibitemOpen
  \bibfield  {author} {\bibinfo {author} {\bibfnamefont {J.}~\bibnamefont
  {Millen}}, \bibinfo {author} {\bibfnamefont {T.}~\bibnamefont {Deesuwan}},
  \bibinfo {author} {\bibfnamefont {P.}~\bibnamefont {Barker}},\ and\ \bibinfo
  {author} {\bibfnamefont {J.}~\bibnamefont {Anders}},\ }\href
  {https://doi.org/10.1038/nnano.2014.82} {\bibfield  {journal} {\bibinfo
  {journal} {Nature Nanotechnology}\ }\textbf {\bibinfo {volume} {9}},\
  \bibinfo {pages} {425} (\bibinfo {year} {2014})}\BibitemShut {NoStop}%
\bibitem [{\citenamefont {Hebestreit}\ \emph {et~al.}(2018)\citenamefont
  {Hebestreit}, \citenamefont {Frimmer}, \citenamefont {Reimann},\ and\
  \citenamefont {Novotny}}]{Hebestreit2018}%
  \BibitemOpen
  \bibfield  {author} {\bibinfo {author} {\bibfnamefont {E.}~\bibnamefont
  {Hebestreit}}, \bibinfo {author} {\bibfnamefont {M.}~\bibnamefont {Frimmer}},
  \bibinfo {author} {\bibfnamefont {R.}~\bibnamefont {Reimann}},\ and\ \bibinfo
  {author} {\bibfnamefont {L.}~\bibnamefont {Novotny}},\ }\href
  {https://doi.org/10.1103/PhysRevLett.121.063602} {\bibfield  {journal}
  {\bibinfo  {journal} {Phys. Rev. Lett.}\ }\textbf {\bibinfo {volume} {121}},\
  \bibinfo {pages} {063602} (\bibinfo {year} {2018})}\BibitemShut {NoStop}%
\bibitem [{\citenamefont {de~los R{\'i}os~Sommer}\ \emph
  {et~al.}(2021)\citenamefont {de~los R{\'i}os~Sommer}, \citenamefont {Meyer},\
  and\ \citenamefont {Quidant}}]{quidant2020}%
  \BibitemOpen
  \bibfield  {author} {\bibinfo {author} {\bibfnamefont {A.}~\bibnamefont
  {de~los R{\'i}os~Sommer}}, \bibinfo {author} {\bibfnamefont {N.}~\bibnamefont
  {Meyer}},\ and\ \bibinfo {author} {\bibfnamefont {R.}~\bibnamefont
  {Quidant}},\ }\href {https://doi.org/10.1038/s41467-020-20419-2} {\bibfield
  {journal} {\bibinfo  {journal} {Nature Communications}\ }\textbf {\bibinfo
  {volume} {12}},\ \bibinfo {pages} {276} (\bibinfo {year} {2021})}\BibitemShut
  {NoStop}%
\bibitem [{\citenamefont {Bowen}\ and\ \citenamefont {Milburn}(2015)}]{bowen}%
  \BibitemOpen
  \bibfield  {author} {\bibinfo {author} {\bibfnamefont {W.}~\bibnamefont
  {Bowen}}\ and\ \bibinfo {author} {\bibfnamefont {G.}~\bibnamefont
  {Milburn}},\ }\href {https://books.google.it/books?id=YZDwCgAAQBAJ} {\emph
  {\bibinfo {title} {Quantum Optomechanics}}}\ (\bibinfo  {publisher} {CRC
  Press},\ \bibinfo {year} {2015})\BibitemShut {NoStop}%
\bibitem [{\citenamefont {Hartmann}\ and\ \citenamefont
  {Plenio}(2008)}]{Hartmann2008}%
  \BibitemOpen
  \bibfield  {author} {\bibinfo {author} {\bibfnamefont {M.~J.}\ \bibnamefont
  {Hartmann}}\ and\ \bibinfo {author} {\bibfnamefont {M.~B.}\ \bibnamefont
  {Plenio}},\ }\href {https://doi.org/10.1103/PhysRevLett.101.200503}
  {\bibfield  {journal} {\bibinfo  {journal} {Phys. Rev. Lett.}\ }\textbf
  {\bibinfo {volume} {101}},\ \bibinfo {pages} {200503} (\bibinfo {year}
  {2008})}\BibitemShut {NoStop}%
\bibitem [{\citenamefont {Romero-Isart}\ \emph {et~al.}(2011)\citenamefont
  {Romero-Isart}, \citenamefont {Pflanzer}, \citenamefont {Blaser},
  \citenamefont {Kaltenbaek}, \citenamefont {Kiesel}, \citenamefont
  {Aspelmeyer},\ and\ \citenamefont {Cirac}}]{Romero2011}%
  \BibitemOpen
  \bibfield  {author} {\bibinfo {author} {\bibfnamefont {O.}~\bibnamefont
  {Romero-Isart}}, \bibinfo {author} {\bibfnamefont {A.~C.}\ \bibnamefont
  {Pflanzer}}, \bibinfo {author} {\bibfnamefont {F.}~\bibnamefont {Blaser}},
  \bibinfo {author} {\bibfnamefont {R.}~\bibnamefont {Kaltenbaek}}, \bibinfo
  {author} {\bibfnamefont {N.}~\bibnamefont {Kiesel}}, \bibinfo {author}
  {\bibfnamefont {M.}~\bibnamefont {Aspelmeyer}},\ and\ \bibinfo {author}
  {\bibfnamefont {J.~I.}\ \bibnamefont {Cirac}},\ }\href
  {https://doi.org/10.1103/PhysRevLett.107.020405} {\bibfield  {journal}
  {\bibinfo  {journal} {Phys. Rev. Lett.}\ }\textbf {\bibinfo {volume} {107}},\
  \bibinfo {pages} {020405} (\bibinfo {year} {2011})}\BibitemShut {NoStop}%
\bibitem [{\citenamefont {Bateman}\ \emph {et~al.}(2014)\citenamefont
  {Bateman}, \citenamefont {Nimmrichter}, \citenamefont {Hornberger},\ and\
  \citenamefont {Ulbricht}}]{Bateman2014}%
  \BibitemOpen
  \bibfield  {author} {\bibinfo {author} {\bibfnamefont {J.}~\bibnamefont
  {Bateman}}, \bibinfo {author} {\bibfnamefont {S.}~\bibnamefont
  {Nimmrichter}}, \bibinfo {author} {\bibfnamefont {K.}~\bibnamefont
  {Hornberger}},\ and\ \bibinfo {author} {\bibfnamefont {H.}~\bibnamefont
  {Ulbricht}},\ }\href {https://doi.org/10.1038/ncomms5788} {\bibfield
  {journal} {\bibinfo  {journal} {Nature Communications}\ }\textbf {\bibinfo
  {volume} {5}},\ \bibinfo {pages} {4788} (\bibinfo {year} {2014})}\BibitemShut
  {NoStop}%
\bibitem [{\citenamefont {Gasbarri}\ \emph {et~al.}(2021)\citenamefont
  {Gasbarri}, \citenamefont {Belenchia}, \citenamefont {Carlesso},
  \citenamefont {Donadi}, \citenamefont {Bassi}, \citenamefont {Kaltenbaek},
  \citenamefont {Paternostro},\ and\ \citenamefont {Ulbricht}}]{Gasbarri2021}%
  \BibitemOpen
  \bibfield  {author} {\bibinfo {author} {\bibfnamefont {G.}~\bibnamefont
  {Gasbarri}}, \bibinfo {author} {\bibfnamefont {A.}~\bibnamefont {Belenchia}},
  \bibinfo {author} {\bibfnamefont {M.}~\bibnamefont {Carlesso}}, \bibinfo
  {author} {\bibfnamefont {S.}~\bibnamefont {Donadi}}, \bibinfo {author}
  {\bibfnamefont {A.}~\bibnamefont {Bassi}}, \bibinfo {author} {\bibfnamefont
  {R.}~\bibnamefont {Kaltenbaek}}, \bibinfo {author} {\bibfnamefont
  {M.}~\bibnamefont {Paternostro}},\ and\ \bibinfo {author} {\bibfnamefont
  {H.}~\bibnamefont {Ulbricht}},\ }\href
  {https://doi.org/10.1038/s42005-021-00656-7} {\bibfield  {journal} {\bibinfo
  {journal} {Communications Physics}\ }\textbf {\bibinfo {volume} {4}},\
  \bibinfo {pages} {155} (\bibinfo {year} {2021})}\BibitemShut {NoStop}%
\end{thebibliography}%

\end{document}